%% file: Data_driven_MFMC_Plasma_phys_final.tex
\documentclass[preprint,12pt,3p]{elsarticle}

\input{settings.tex}

\journal{J.~Comp.~Phys.}

\begin{document}

\begin{frontmatter}

\title{Data-driven low-fidelity models for multi-fidelity Monte Carlo sampling in plasma micro-turbulence analysis}
\author[label1, label0]{Julia Konrad}
\address[label1]{Department of Informatics, Technical University of Munich}
\ead{julia.konrad@tum.de}

\author[label2, label0]{Ionu\cb{t}-Gabriel Farca\cb{s}}
\address[label2]{Oden Institute for Computational Engineering and Sciences, The University of Texas at Austin}
\ead{ionut.farcas@austin.utexas.edu}

\fntext[label0]{These authors contributed equally}

\author[label3]{Benjamin Peherstorfer}
\address[label3]{Courant Institute of Mathematical Sciences, New York University}
\ead{pehersto@cims.nyu.edu}

\author[label2]{Alessandro Di Siena}
\ead{alessandro.disiena@austin.utexas.edu}

\author[label1,label2,label4]{Frank Jenko}
\address[label4]{Max Planck Institute for Plasma Physics}
\ead{frank.jenko@ipp.mpg.de}

\author[label1]{Tobias Neckel}
\ead{neckel@in.tum.de}

\author[label1]{Hans-Joachim Bungartz}
\ead{bungartz@in.tum.de}

\input{sections/abstract}  


\end{frontmatter}

\input{sections/01_introduction}
\input{sections/02_preliminaries}
\input{sections/03_models}
\input{sections/04_numerical_results}

\input{sections/05_conclusions}
\input{sections/acknowledgements}

\bibliographystyle{elsarticle-harv}
\bibliography{Data_driven_MFMC_Plasma_phys_final}

\end{document}

%% file: settings.tex
\newcount\Comments
\usepackage[utf8]{inputenc}
\usepackage{amssymb}
\usepackage{amsfonts}
\usepackage{graphicx}
\usepackage{multirow}
\usepackage{bigstrut}
\usepackage{epstopdf}
\usepackage[english]{babel}
\usepackage{combelow}
\usepackage{amsmath}
\usepackage{amsthm}
\usepackage{algorithm}
\usepackage{algorithmicx}
\usepackage[noend]{algpseudocode}
\usepackage[dvipsnames]{xcolor}
\usepackage{bm}
\usepackage{todonotes}
\usepackage{tikz}
\usetikzlibrary{shapes}
\usetikzlibrary{arrows}
\usetikzlibrary{positioning}
\usetikzlibrary{matrix}
\usetikzlibrary{patterns}
\usetikzlibrary{decorations.pathreplacing,decorations.pathmorphing,decorations.markings}
\usepackage{tikz-qtree}
\usetikzlibrary{trees,calc,arrows.meta,positioning,bending}
\usepackage{mathtools}
\usepackage{todonotes}
\usepackage[version=3]{mhchem}

\newdefinition{remark}{Remark}
\numberwithin{equation}{section}

\newcommand{\R}{\mathbb{R}}
\newcommand{\E}{\mathbb{E}}
\newcommand{\Var}{\textrm{Var}}

\newcommand{\Xcal}{\mathcal{X}}
\newcommand{\Ycal}{\mathcal{Y}}
\newcommand{\Ucal}{\mathcal{U}}
\newcommand{\bftheta}{\boldsymbol\theta}
\newcommand{\bfTheta}{\boldsymbol\Theta}

\newcommand{\kibitz}[2]{\ifnum\Comments=1\textcolor{#1}{#2}\fi}

\definecolor{brandeisblue}{rgb}{0.0, 0.44, 1.0}

\makeatletter
\newcommand*{\centerfloat}{%
  \parindent \z@
  \leftskip \z@ \@plus 1fil \@minus \textwidth
  \rightskip\leftskip
  \parfillskip \z@skip}
\makeatother

%% file: sections/abstract.tex
\begin{abstract}
The linear micro-instabilities driving turbulent transport in magnetized fusion plasmas (as well as the respective nonlinear saturation mechanisms) are known to be sensitive with respect to various physical  parameters characterizing the background plasma and the magnetic equilibrium.
Therefore, uncertainty quantification is essential for achieving predictive numerical simulations of plasma turbulence.
However, the high computational costs of the required gyrokinetic simulations and the large number of parameters render standard Monte Carlo techniques intractable.
To address this problem, we propose a multi-fidelity Monte Carlo approach in which we employ data-driven low-fidelity models that exploit the structure of the underlying problem such as low intrinsic dimension and anisotropic coupling of the stochastic inputs.
The low-fidelity models are efficiently constructed via sensitivity-driven dimension-adaptive sparse grid interpolation using both the full set of uncertain inputs and subsets comprising only selected, important parameters.
We illustrate the power of this method by applying it to two plasma turbulence problems with up to $14$ stochastic parameters, demonstrating that it is up to four orders of magnitude more efficient than standard Monte Carlo methods measured in single-core performance, which translates into a runtime reduction from around eight days to one hour on 240 cores on parallel machines.
\end{abstract}

%% file: sections/01_introduction.tex
\section{Introduction}\label{sec:intro}

The goal of creating ``burning" plasmas in a toroidal magnetic device represents a milestone on the way towards the practical realization of fusion power on Earth.
Efforts are currently underway to take this important step in the international ITER experiment, with operation projected to begin within the next few years. 
One of the key physics problems in this context is the understanding, prediction, and control of plasma turbulence.
Turbulent fluctuations of various plasma quantities induce, in particular, high levels of cross-field heat transport, thus determining the so-called energy confinement time.
The latter needs to exceed a certain threshold (according to the Lawson criterion) to achieve a burning plasma state.
The linear micro-instabilities driving this turbulent transport (as well as the respective nonlinear saturation mechanisms) are known to be very sensitive with respect to various physical parameters characterizing the background plasma and the magnetic equilibrium.
Moreover, experimental measurements of these parameters are, of course, subject to uncertainties.
Therefore, the development of a predictive capability calls for a systematic quantification of uncertainties.
However, the high computational costs of numerical simulations and the potentially large number of stochastic inputs, parameters, and coefficients make uncertainty quantification challenging for these problems \cite{HJ17}.

\subsection{Towards quantifying uncertainty in plasma turbulence problems}
To overcome the challenges of quantifying uncertainty in real-world problems such as plasma micro-turbulence simulations, we present a framework based on multi-fidelity Monte Carlo (MFMC) sampling \cite{NME:NME4761,PWK16MFMCAsymptotics,PWG16}.
MFMC reduces the mean-squared error of standard Monte Carlo (MC) estimators by making use of control variates \cite{Ne87}.
Specifically, in MFMC, the control variates are standard MC estimators given by low-fidelity models, which are computationally cheap approximations of the underlying high-fidelity model.
Examples of low-fidelity models include reduced-physics models, e.g., coarse-grid approximations,  data-fit low-fidelity models, such as interpolation, spectral projection or regression, or projection-based reduced models, including the proper orthogonal decomposition or the reduced basis method; see \cite{PWG18} and the references therein for more details.

Control variates have been used to reduce the error of MC estimators in kinetic models such as Boltzmann's equation in \cite{DP19}.
The work \cite{DP19} was generalized in \cite{DP20}, where multiple control variates are employed.
In addition, in \cite{Ko19}, MFMC was employed in a benchmark scenario prominent in plasma micro-turbulence simulations \cite{Di00}.
In \cite{Fa20b, P19AMFMC}, a context-aware version of MFMC is formulated, in which data-driven low-fidelity models are explicitly constructed to being used together with the high-fidelity model; see also \cite{AP20Context}. 

Besides sampling-based methods, stochastic collocation and quasi-Monte Carlo have also been used to address the challenges of uncertainty  in plasma physics.
In \cite{La20}, generalized polynomial chaos and quasi-Monte Carlo methods were employed in multiscale plasma fusion simulations.
Moreover, \cite{VH18} used nonintrusive stochastic collocation methods based on non-uniform interpolation sequences in a drift-wave turbulence study from a linear plasma experiment.
In addition, adaptive sparse grids \cite{BG04} were considered in \cite{Ya20} to compute the runaway probability of electrons.

\subsection{Novelty and contribution of proposed data-driven multi-fidelity approach for uncertainty propagation in plasma turbulence problems}
In this work, we employ the well established plasma turbulence code {\sc Gene} \cite{Je00} which solves the gyrokinetic Vlasov-Maxwell coupled system of partial differential equations in $5$D state space, focusing on calculating the linear growth rates of the dominant micro-instabilities.
For constructing low-fidelity models, we primarily build on approximations with adaptive sparse grids that have been shown to be effective in quantifying uncertainties in complex applications:
In \cite{Fa20, Fa20b}, a sensitivity-driven dimension-adaptive sparse grid approximation method was introduced to efficiently compute quantities of interest such as expectation, variance and Sobol' indices for global sensitivity analysis \cite{So01}.
The sensitivity-driven approach exploits the fact that typically in most real-world problems (i) the intrinsic stochastic dimension is lower than the ambient dimension and (ii) the uncertain inputs are anisotropically coupled.
In \cite{Fa20, Fa20b}, the sensitivity-driven approach was applied to two plasma micro-turbulence test cases, including a scenario with $12$ uncertain inputs based on an ASDEX Upgrade discharge \cite{Fr18}.
The same approach was recently used in a study concerning turbulence stabilization via energetic particles with $21$ stochastic parameters \cite{FDJ21}.

In our MFMC approach, we leverage the sensitivity-driven sparse-grid models as low-fidelity models.
Furthermore, we exploit sensitivity information to reduce the stochastic dimension. This has also been studied in \cite{Fa18} in the context of multilevel stochastic collocation for fluid-structure interaction simulations.
In this paper, we rely on the feature that the construction of sparse-grid models reveals detailed information about the sensitivity of each stochastic input and use this information to additionally construct low-fidelity models with lower input dimensions, i.e., models in which only few of the inputs are varied while the remaining inputs are fixed to their mean value.

To summarize, the present paper is novel in at least two ways: from a methodological perspective, we employ data-driven, structure-exploiting low-fidelity models in MFMC with the goal to make MFMC feasible in computationally expensive, real-world problems for which standard sampling methods are unfeasible.
From an application scenario point of view, we employ our MFMC approach in large-scale plasma micro-instability analysis, a problem of high practical relevance.

\subsection{Outline of this paper}
This paper is organized as follows.
In Section \ref{sec:background}, we explain the notation and summarize the foundations of multi-fidelity uncertainty propagation.
In Section \ref{sec:data_driven_surrogates}, we describe the hierarchy of models used in this paper.
Our numerical results in two plasma micro-turbulence scenarios are presented in Section \ref{sec:results}.  
The first scenario is a modified benchmark in which we consider either three or eight uncertain parameters (Section \ref{subsec:cbc}). 
The second scenario discussed in Section \ref{subsec:jet} is concerned with turbulence stabilization via energetic particles with $14$ uncertain inputs, for which one realization of the stochastic inputs requires $240$ cores and a total runtime of about $3.2$ hours.
We close with conclusions in Section~\ref{sec:conclusions}.

%% file: sections/02_preliminaries.tex
\section{Multi-fidelity uncertainty propagation} \label{sec:background}

In this section, we introduce notation and provide a brief summary of the mathematical foundations of multi-fidelity uncertainty propagation with MFMC.

\subsection{Uncertainty propagation} \label{subsec:prob_statement}

Let $f^{(0)}: \Xcal \to \Ycal$ be a bounded and measurable function with respect to the Lebesgue measure and the Borel $\sigma$-algebra on $\mathbb{R}$. 
The domain $\Xcal \subset \mathbb{R}^d$ is the set of $d$-dimensional inputs $\boldsymbol{\theta} = [\theta_1, \theta_2, \ldots, \theta_d]^T$ and the domain $\Ycal \subset \mathbb{R}$ is the range of scalar-valued outputs $y = f^{(0)}(\bftheta)$. 
In the following, we consider the situation where $f^{(0)}$ is computationally expensive to evaluate because evaluating $f^{(0)}$ entails large-scale, high-fidelity numerical simulations.
Denote the computational costs of evaluating $f^{(0)}$ at an input $\bftheta \in \Xcal$ as $w_0 \gg 0$.

Let now $\bfTheta = [\Theta_1, \dots, \Theta_d]^T$ be a random vector with image $\Xcal$ and with probability density function $\pi$. 
The goal of uncertainty propagation is to estimate quantities of interest such as the expected value of the output random variable $f^{(0)}(\bfTheta)$
\begin{equation} \label{eq:hi_fi_expec}
    \mu_0 = \E[f^{(0)}(\bfTheta)] = \int_{\Xcal} f^{(0)}(\boldsymbol{\theta}) \pi(\boldsymbol{\theta}) \mathrm{d}\boldsymbol{\theta}
\end{equation}
and its variance 
\begin{equation} \label{eq:hi_fi_var}
    \sigma^2_0 = \Var[f^{(0)}(\bfTheta)] = \E[\left(f^{(0)}(\bfTheta)\right)^2] - \E[f^{(0)}(\bfTheta)]^2.
\end{equation}
MC estimators of the expected value and variance of  $f^{(0)}(\bfTheta)$ are
\begin{equation}\label{eq:mean_var_mc}
    \hat{E}_n^{(0)} := \frac{1}{n} \sum_{i=1}^n f^{(0)}(\boldsymbol{\theta}_i), \quad 
    \widehat{\Var}_n^{(0)} := \frac{1}{n-1} \sum_{i=1}^n (f^{(0)}(\boldsymbol{\theta}_i) - \hat{E}_n^{(0)})^2,
\end{equation}
respectively, where $\boldsymbol{\theta}_1, \boldsymbol{\theta}_2, \ldots, \boldsymbol{\theta}_n$ are $n$ independent and identically distributed (i.i.d.) samples from $\bfTheta$. 
The costs of the MC estimator $\hat{E}_n^{(0)}$ of $\mathbb{E}[f^{(0)}(\bfTheta)]$ are $p = nw_0$ because $f^{(0)}$ is evaluated at $n$ realizations. 
The mean-squared error (MSE) of the estimator $\hat{E}_n^{(0)}$ is
\begin{equation}\label{eq:mse_mc}
    \mathrm{MSE}[\hat{E}_n^{(0)}] = \frac{\sigma_0^2}{p} w_0.
\end{equation}
Because we consider the situation in which the evaluation costs of $f^{(0)}$ are large, the slow decay of the MSE in \eqref{eq:mse_mc} with respect to the budget $p$ makes the MC estimator $\hat{E}_n^{(0)}$ computationally prohibitive.
For example, in Section \ref{subsec:jet}, one plasma micro-turbulence simulation requires, on average, a $w_0$ of about $3.2$ hours on 240 cores in total.
A large number of such simulations is not feasible.

\subsection{Multi-fidelity Monte Carlo sampling}\label{subsec:mfmc}

The MFMC method proposed in \cite{NME:NME4761,PWG16} exploits that in many problems in science and engineering we either have available or can construct low-fidelity models that approximate $f^{(0)}$ with lower computational costs than $w_0$.
Let us assume we have available $k$ low-fidelity models $f^{(1)}$, $f^{(2)}$, \ldots, $f^{(k)}$.
MFMC exploits the lower computational cost of the low-fidelity models to obtain an estimator that has lower costs than a standard MC estimator with the same MSE.
MFMC guarantees unbiasedness even if the approximation error of the low-fidelity models cannot be bounded rigorously.
The quality of low-fidelity models for MFMC is ascertained via their Pearson correlation coefficient with respect to $f^{(0)}$,
\begin{equation*}
\rho_{j} = \frac{\mathrm{Cov[f^{(0)}, f^{(j)}]}}{\sigma_0\sigma_j} \in [-1, 1], \quad j = 1, 2, \ldots, k,
\end{equation*}
where $\sigma_j^2$ denotes the variance of model $j$ for $j=0, 1, \ldots, k$ and $\operatorname{Cov}$ is the covariance.

Let $m_j \in \mathbb{N}$ for $i = 0, 1, \ldots, k$ denote the number of evaluations of model $f^{(j)}$ such that
\begin{equation}\label{eq:ascending_m_j}
1 \leq m_0 < m_1 < \ldots < m_k.
\end{equation}
That is, we ensure that the high-fidelity model is evaluated at least once.
Consider $m_k$ i.i.d.~realizations drawn from the input density, $\pi(\boldsymbol{\theta})$:
\begin{equation*}
    \boldsymbol{\theta}_1, \boldsymbol{\theta}_2, \ldots, \boldsymbol{\theta}_{m_k}.
\end{equation*}
Model $f^{(j)}$ for $j = 0, 1, \ldots, k$ is evaluated at the first $m_j$ samples $\boldsymbol{\theta}_1, \boldsymbol{\theta}_2, \ldots, \boldsymbol{\theta}_{m_j}$ to obtain 
\begin{equation}
    f^{(j)}(\boldsymbol{\theta}_1), f^{(j)}(\boldsymbol{\theta}_2), \ldots, f^{(j)}(\boldsymbol{\theta}_{m_j}).
\end{equation}
We compute the following standard MC estimators for expectation and variance:
\begin{equation}\label{eq:est_m_j}
    \hat{E}_{m_j}^{(j)} = \frac{1}{m_j} \sum_{m=1}^{m_j} f^{(j)}(\boldsymbol{\theta}_m) 
    , \quad
    \widehat{\Var}_{m_j}^{(j)} = \frac{1}{m_j - 1} \sum_{m=1}^{m_j} (f^{(j)}(\boldsymbol{\theta}_m) - \hat{E}_{m_j}^{(j)})^2.
\end{equation}
Additionally, for the low-fidelity models $f^{(j)}$ with $j = 1, 2, \ldots, k$ we compute the estimators
\begin{equation}\label{eq:est_m_j-1}
    \hat{E}_{m_{j - 1}}^{(j)} = \frac{1}{m_{j - 1}} \sum_{m=1}^{m_{j - 1}} f^{(j)}(\boldsymbol{\theta}_m) 
    , \quad
    \widehat{\Var}_{m_{j - 1}}^{(j)} = \frac{1}{m_{j - 1} - 1} \sum_{m=1}^{m_{j - 1}} (f^{(j)}(\boldsymbol{\theta}_m) - \hat{E}_{m_{j - 1}}^{(j)})^2.
\end{equation}
Note that the estimators in \eqref{eq:est_m_j} use the full set of evaluations of model $f^{(j)}$, while the estimators in \eqref{eq:est_m_j-1} use the first $m_{j - 1}$ evaluations.
This makes the estimators in \eqref{eq:est_m_j-1} dependent on the estimators in \eqref{eq:est_m_j}.

The MFMC estimator for the expectation of the high-fidelity model is given by
\begin{equation}\label{eq:mean_mfmc}
    \hat{E}^{\mathrm{MFMC}} := \hat{E}_{m_0}^{(0)} + \sum_{j=1}^k \alpha_j (\hat{E}_{m_j}^{(j)} - \hat{E}_{m_{j-1}}^{(j)})
\end{equation}
and the estimator for the variance is
\begin{equation}\label{eq:var_mfmc}
    \widehat{\Var}^{\mathrm{MFMC}} := \widehat{\Var}_{m_0}^{(0)} + \sum_{j=1}^k \alpha_j (\widehat{\Var}_{m_j}^{(j)} - \widehat{\Var}_{m_{j-1}}^{(j)})
\end{equation}
where $\alpha_1, \alpha_2, \ldots, \alpha_k \in \R$.

Let $w_j$ denote the runtime of the low-fidelity model $f^{(j)}$ for $j = 1, 2, \ldots, [w_0, w_1, \ldots, w_k]^T$.
Given a budget $p := \sum_{j=0}^k w_j m_j$, in \cite[Theorem 3.4]{PWG16} it is shown that if
\begin{equation} \label{eq:MFMC_assump}
1 =: \rho_0 > \rho_1 > \rho_2 > \ldots > \rho_k  \text{ and } \frac{w_{j - 1}}{w_j} > \frac{\rho_{j-1}^2 - \rho_{j}^2}{\rho_{j}^2 - \rho_{j + 1}^2}, \quad j = 1, 2, \ldots, k, 
\end{equation}
then the MSE of the MFMC mean estimator \eqref{eq:mean_mfmc} is minimized given the budget $p$, and
\begin{equation} \label{eq:MFMC_opt_params}
\alpha_j^* = \frac{\rho_j \sigma_1}{\sigma_j}, \quad m_0^* = \frac{p}{\boldsymbol{w}^T \boldsymbol{r}^*}, \quad m_j^* = r_j^* m_0^*, \quad j = 1, 2, \ldots, k,
\end{equation}
where $\boldsymbol{r}^* := [r_0^*, r_1^*, \ldots, r_k^*]^T, r_j^* = \sqrt{w_0(\rho_j^2 - \rho_{j + 1}^2)/w_j(1 - \rho_1^2)}$.
Using the optimal number of model evaluations $m_0^*, m_1^*, \ldots, m_k^*$ and coefficients $\alpha_1^*, \alpha_2^*, \ldots, \alpha_k^*$, the MSE of $\hat{E}^{\mathrm{MFMC}}$ is
\begin{equation}\label{eq:mse_mfmc}
\mathrm{MSE}[\hat{E}^{\mathrm{MFMC}}] = \frac{\sigma_0^2}{p} \left( \sum_{j=0}^k \sqrt{w_j (\rho_{j}^2 - \rho_{j+1}^2)} \right) ^2.
\end{equation}
If we compare the above MSE with the MSE of the standard MC mean estimator \eqref{eq:mse_mc}, it holds that $\mathrm{MSE}[\hat{E}^{\mathrm{MFMC}}] <\mathrm{MSE}[\hat{E}^{\mathrm{MC}}]$ if and only if
\begin{equation}\label{eq:var_reduction}
   \gamma \coloneqq \left( \sum_{j=0}^k \sqrt{\frac{w_j}{w_0} (\rho_{j}^2 - \rho_{j+1}^2)} \right) ^2 < 1.
\end{equation}
The above formula gives a means to determine which low-fidelity models lead to an MFMC estimator with a smaller MSE than the MSE of the standard MC estimator for which the same budget, $p$, was used.
In \cite[Algorithm 1]{PWG16}, a model selection algorithm is proposed, which, given a set of low-fidelity models, selects the subset that leads to the MFMC estimator with the lowest MSE.

In the specific situation of two models and that the expected value of the surrogate-model output $f^{(1)}(\theta)$ is known in closed form, it can be used in \eqref{eq:mean_mfmc} directly instead of the estimator $\hat{E}_{m_1}^{(1)}$. This specific situation is investigated with control functionals in \cite{https://doi.org/10.1111/rssb.12185} and in the context of MFMC in the asymptotic limit in \cite{P19AMFMC}. However, for typical surrogate models, a closed form of the expected value is unavailable and one has to rely on the estimator $\hat{E}_{m_1}^{(1)}$.

We point out that the standard MFMC algorithm of \cite{PWG16} is for the estimation of the mean of the high-fidelity model.
That is, given a budget $p$ and assuming that \eqref{eq:MFMC_assump} holds true, the minimization of the MSE of the mean estimator leads to the optimal parameters in \eqref{eq:MFMC_opt_params}.
These parameters are not necessarily optimal for estimating the variance of the high-fidelity model with the MFMC estimator \eqref{eq:var_mfmc}.
Nevertheless, in \cite{Qi18}, it was shown that using the mean-optimal parameters for estimating the variance too leads to a comparable MSE as when directly minimizing the MSE of the variance estimator, which is significantly more challenging since the variance is a nonlinear operator.
Thus, the realizations of the model outputs can be used to compute both mean and variance estimators.

\begin{remark} \label{remark:remark_rho_cost_computation}
The ingredients in MFMC are the correlation coefficients, $\rho_j$, and evaluation costs, $w_j$, of the models in the hierarchy.
In some problems the quantities are readily available from physical and theoretical insights.
If the quantities are unavailable, then they can estimated from typically a small number of pilot samples \cite{PWG16}. We will demonstrate with our numerical examples in Section~\ref{subsec:jet} that the pilot samples can be re-used in the MFMC estimator. Re-using the pilot samples can introduce a bias; however, the mean-squared error of the estimator is typically dominated by the limited number of samples rather than the bias due to re-using pilot samples.  
\end{remark}

%% file: sections/03_models.tex
\section{Plasma micro-turbulence models: high-fidelity and data-driven low-fidelity models}\label{sec:data_driven_surrogates}

In this section, we present the hierarchy of models used in our numerical experiments.
In Section \ref{subsec:plasma}, we summarize the gyrokinetic theory for plasma micro-turbulence simulations, which is the basis for the high-fidelity model employed in this work.
The main methodology used to construct data-driven low-fidelity models via sensitivity-driven sparse grids is presented in Section \ref{subsec:sg}.
For a broader overview, we also consider a second type of data-driven low-fidelity models based on deep neural networks, which we summarize in Section \ref{subsec:ml}.

\subsection{High-fidelity model}\label{subsec:plasma}

We build on gyrokinetic theory \cite{BH07} to assess plasma micro-turbulence, which reduces the dimension of the state space of classical kinetic models from six (three positions, three velocities) to five (three positions, two velocities). At the same time, a number of small---and irrelevant (for the specific scenario at hand) ---space-time scales are removed from the problem, leading to savings of several orders of magnitude in terms of computational costs. 
Throughout this paper the gyrokinetic equations are solved numerically with the simulation code {\sc Gene} \cite{Je00}.
In the gyrokinetic code {\sc Gene} each particle species, $s$, is characterized by a distribution function, which is split into a static background ($\pi_0$) -- often assumed a Maxwellian distribution function --  and a fluctuating part ($\pi_1$)
\begin{equation*}
\pi_s(t, \boldsymbol{R}, v_{||}, \mu_m) = \pi_{0,s}(\boldsymbol{R}, v_{||}, \mu_m) + \pi_{1,s}(t, \boldsymbol{R}, v_{||}, \mu_m),
\end{equation*}
where $t$ is time, $\boldsymbol{R} = (R_x, R_y, R_z)$ describes the position of the gyrocenter in real space, $v_{||}$ denotes the velocity parallel to the background magnetic field, and $\mu_m := m_s v_{\perp}^2/2B$ is the magnetic moment, where $m_s$ denotes the mass of species $s$, $v_{\perp}$ is the perpendicular component of the velocity and $B$ denotes the magnetic field. Here, it is assumed that $\pi_1 / \pi_0 \sim \epsilon_{\delta} \ll 1$. 
The dynamics of the perturbed part of the distribution function $\pi_{1,s}$ are modelled by the first order (in the expansion parameter $\epsilon_{\delta}$) gyrokinetic Vlasov equation with collisions:
\begin{equation}\label{eq:vlasov}
\begin{split}
C(\pi_s, \pi_{s'}) & = \dot{\pi}_{1,s} + \boldsymbol{v_{1,c}} \cdot \left(\boldsymbol{\nabla} \pi_{0,s} - \frac{\mu_m}{m_s v_{||}} \frac{\partial \pi_{0,s}}{\partial v_{||}}\boldsymbol{\nabla}B_0\right) + \left(v_{||} \boldsymbol{\hat{b}_0} + \boldsymbol{v_{1,c}} \right) \cdot \\
& \left( \boldsymbol{\nabla} \pi_{1,s} + \frac{1}{m_s v_{||}} \left( q_s \boldsymbol{E}_1 - \mu_m \boldsymbol{\nabla} B_{1,||}\right) \frac{\partial \pi_{0,s}}{\partial v_{||}} + \frac{1}{m_s v_{||}} \left( q_s \boldsymbol{E}_1 - \mu_m \boldsymbol{\nabla} \left( B_0 + B_{1,||}\right)\right) \frac{\partial \pi_{1,s}}{\partial v_{||}} \right),
\end{split}
\end{equation}
where $C(\pi_s, \pi_{s'})$ is a collision operator, $B_0$ the background magnetic field, $\boldsymbol{E}_1$ the perturbed part of the electric field and $\boldsymbol{v_{1,c}}$ the drift velocity due to the fields. 
To self-consistently evolve the perturbed part of the distribution function $\pi_{1,s}$,  we also have to advance three scalar quantities characterizing the electromagnetic fields by solving Maxwell’s equations: the electrostatic potential, $\Phi_1$, the parallel component of the vector potential, $A_{1, ||}$, which is linked to the perpendicular magnetic field perturbations, and the parallel component of the magnetic field perturbations, $B_{1, ||}$, i.e.,
\begin{equation}
\begin{split}
\text{Poisson's \ equation: \ \ } & \boldsymbol{\nabla} \cdot \boldsymbol{E_1} = -\nabla^2 \Phi_1 = 4\pi \sum_s q_s n_{1s}, \\
\text{Amp\`ere's \ law for $A_{1,||}$: \ \ } & - \nabla^2 A_{1,||} = \frac{4 \pi}{c} \sum_s n_{1s} q_s u_{1s,||}, \\
\text{Amp\`ere's \ law for $B_{1,||}$: \ \ } & \left(\boldsymbol{\nabla} \times \boldsymbol{B}_1\right)_\perp = \frac{4 \pi}{c} \sum_s n_{1s} \boldsymbol{u}_{1s,\perp},
\end{split}
\label{eq:fields}
\end{equation}
where $q_s$ is the charge, $c$ is the speed of light, and $n_{1s}$ is the $0$th space moment, $u_{1s,||}$ the $1$st order velocity moment in $v_{||}$ of $\pi_s$ and $\boldsymbol{u}_{1s,\perp}$ the $1$st order velocity moment in $\mu_m$ of $\pi_s$. 
The velocity integrals of $\pi_1$ are computed in the particle coordinates, while the Vlasov equation is written in the gyrocenter coordinates \cite{Dannert_2005}. 
The system of equations \eqref{eq:vlasov}-\eqref{eq:fields} defines the partial differential Vlasov-Maxwell equations solved in {\sc Gene}.

In {\sc Gene}, the gyrokinetic Vlasov-Maxwell system of nonlinear PDEs is solved using the method of lines: the spatial operators are discretized using either finite differences on equidistant points or spectral (Fourier) decompositions, while the resulting system of ODEs is efficiently integrated using explicit methods, such as fourth-order Runge-Kutta. 

In this work, we are interested in linear (in phase space variables $\boldsymbol{R}, v_{||}, \mu_m$) gyrokinetic simulations, which can be used to characterize the underlying micro-instabilities. 
The equations employed in linear simulations are obtained from the gyrokinetic Vlasov-Maxwell system \eqref{eq:vlasov}-\eqref{eq:fields} by simply neglecting all nonlinear terms.
In these simulations, {\sc Gene} is run in the so-called flux-tube limit, which allows having periodic boundary conditions over the radial ($x$) direction \cite{Dannert_2005}. 
The generic form of linear gyrokinetic simulations reads:
\begin{equation*}
\dot{\boldsymbol{\pi}}_s = \mathbb{O}_{\mathrm{lin}}(\boldsymbol{\pi}_s),
\end{equation*}
where $\boldsymbol{\pi}_s$ is a vector holding the discretized five-dimensional distribution function of species $s$.
The discrete form of the operator $\mathbb{O}_{\mathrm{lin}}$ is a matrix, $O_{\mathrm{lin}}$, which leads to
\begin{equation*}
\dot{\boldsymbol{\pi}}_s = O_{\mathrm{lin}} \boldsymbol{\pi}_s.
\end{equation*}
To obtain the above linear equation, {\sc Gene} employs Fourier decompositions in the radial ($x$) direction and equidistant points along the field line, i.e., in the $z$ direction; in flux-tube simulations, only one point is used in the bi-normal ($ky$) direction. 
In velocity space, equidistant points are applied for the parallel velocity, while Gauss-Laguerre quadrature points are used to discretize the magnetic moment.

The high-fidelity model used in this work computes the growth rate or spectral abscissa, i.e., the maximum real part over the spectrum, which we denote by $\gamma_1$:
\begin{equation}
f^{(0)}(\bfTheta) = \gamma_1 (\bfTheta)
\end{equation}
which characterizes the dominant micro-instability mode.
The entries in $\bfTheta$ are parameters characterizing the particle species, such as temperature, density and their (logarithmic) gradients, parameters characterizing the magnetic geometry or parameters associated to electromagnetic effects, if electrons are used in the gyrokinetic equations etc.
In this work, $\bfTheta$ has up to $14$ components, including the safety factor, $q$, which describes the relationship of the number of toroidal turns of a magnetic field line to number of poloidal turns; and its derivative, the magnetic shear, $\hat{s}$; $\beta$, the ratio of kinetic to magnetic pressure inside the plasma; the normalized collision frequency, $\nu_c$; and parameters characterizing the underlying species, such as their density, $n_{s}$, and its negative logarithmic gradient, $\omega_{n_{s}}$, and their temperature, $T_{s}$, and its negative logarithmic gradient, $\omega_{T_{s}}$.
For more details about the gyrokinetic equations solved by {\sc Gene}, we refer the reader to \cite{Goe11} and the references therein.

\subsection{Low-fidelity model: Sensitivity-driven dimension-adaptive sparse grid interpolation}\label{subsec:sg}

In this section, we summarize the sensitivity-driven dimension-adaptive sparse grid interpolation approach formulated in \cite{Fa20,Fa20b}, which is our main methodology used to create data-driven low-fidelity models for MFMC.
Our notation is similar to \cite{Fa20,Fa20b}.

\subsubsection{Interpolation on sparse grids}
The sparse grid interpolation \cite{BG04} low-fidelity model reads
\begin{equation} \label{eq:tensor_delta_finite}
\Ucal^{d}_{\mathcal{L}}[f^{(0)}] = \sum_{\boldsymbol{\ell} \in \mathcal{L}} \Delta^{d}_{\boldsymbol{\ell}}[f^{(0)}],
\end{equation}
where $\boldsymbol{\ell} = (\ell_1, \ell_2, \ldots \ell_{d}) \in \mathbb{N}^{d}$ denotes a multiindex, $\mathcal{L} \subset \mathbb{N}^d$ is a multiindex set, and
\begin{equation} \label{eq:hierarch_surplus}
\Delta^{d}_{\boldsymbol{\ell}}[f^{(0)}] = \sum_{\boldsymbol{z} \in \{0, 1\}^{d}} (-1)^{|\boldsymbol{z}|_1} \Ucal^{d}_{\boldsymbol{\ell} - \boldsymbol{z}}[f^{(0)}]
\end{equation}
are the so-called hierarchical surpluses, where $|\boldsymbol{z}|_1 := \sum_{i=1}^{d} z_i$.
The surpluses are computed from full-grid operators, $\Ucal^{d}_{\boldsymbol{\ell}}$, which are tensorizations of one-dimensional approximations:
\begin{equation} \label{eq:tensor_1D}
\Ucal^{d}_{\boldsymbol{\ell}}[f^{(0)}] = \left(\bigotimes_{i=1}^{d} \Ucal^{i}_{\ell_i}\right)[f^{(0)}].
\end{equation}
Therefore, the sparse grid interpolation low-fidelity model \eqref{eq:tensor_delta_finite} is a linear combination of hierarchical surpluses \eqref{eq:hierarch_surplus}, which are computed from tensorizations \eqref{eq:tensor_1D} of one-dimensional operators, $\Ucal^{i}_{\ell_i}$.
To ensure that \eqref{eq:tensor_1D} can be computed, we need the following two assumptions: the image, $\Xcal$, of the stochastic inputs needs to have a product structure:
\begin{equation*}
\Xcal := \bigotimes_{i=1}^d \mathcal{X}_i,
\end{equation*}
which implies that the input density, $\pi$, needs to have a product structure as well:
\begin{equation*}
\pi(\boldsymbol{\theta}) := \prod_{i=1}^d \pi_i(\theta_i),
\end{equation*}
where $\mathcal{X}_i$ is the image of the density $\pi_i$ associated with input $\theta_i$.
The latter assumptions means that the stochastic inputs need to be independent.
This can be relaxed if a (possibly nonlinear) transformation is used, e.g., a transport map \cite{Ma16}. 

In this work, $\Ucal^{i}_{\ell_i}$ are interpolation operators based on Lagrange polynomials (see, e.g., \cite{BT04, CEP12, Fa20, Fa20b, NTW08}), constructed using weighted (L)-Leja points \cite{JWZ18, NJ14}.
Let $g : \mathcal{X}_i \rightarrow \mathbb{R}$ be a univariate function and let $\mathbb{P}_{P_{\ell_i}}$ be the space of univariate polynomials of degree $P_{\ell_i} \in \mathbb{N}$ for $i =1, 2, \ldots, d$. 
Univariate Lagrange interpolation is defined as:
\begin{equation} \label{eq:interp_1D}
\Ucal_{\ell_i}^{i} : C^0(\mathcal{X}_i) \rightarrow \mathbb{P}_{P_{\ell_i}}, \quad \Ucal_{\ell_i}^{i}[g] := \sum_{n = 1} ^ {\ell_i} g(\theta_n)L_n(\theta),
\end{equation}
where $\{\theta_n\}_{n=1}^{\ell_i}$ are weighted (L)-Leja points computed w.r.t.~the density $\pi_i$:
\begin{equation*}
\begin{split}
& \theta_1 = \underset{\theta \in \mathcal{X}_i}{\mathrm{argmax}} \ \pi_i(\theta) \\
\ & \theta_n = \underset{\theta \in \mathcal{X}_i}{\mathrm{argmax}} \ \pi_i(\theta)\prod_{m=1}^{n-1} \left| (\theta - \theta_m) \right|, \quad n = 2, 3, \ldots, \ell_i,
\end{split}
\end{equation*}
and $\{L_n(\theta)\}_{n=1}^{\ell_i}$ are Lagrange polynomials of degree $n - 1$ satisfying the interpolation condition $L_n(\theta_m) = \delta_{nm}$, where $\delta_{nm}$ is Kronecker's delta function. 
For improved numerical stability, we implement \eqref{eq:interp_1D} in terms of the barycentric formula (see, e.g., \cite{BT04}).
It follows that the multivariate interpolation operator \eqref{eq:tensor_1D} is determined as
\begin{equation} \label{eq:interp_ND}
\Ucal^{d}_{\boldsymbol{\ell}}[f^{(0)}] = \sum_{\boldsymbol{p} \in \mathcal{P}_{\boldsymbol{\ell}}} f^{(0)}(\boldsymbol{\theta}_{\boldsymbol{p}}) L_{\boldsymbol{p}}(\boldsymbol{\theta}),
\end{equation}
where $L_{\boldsymbol{p}}(\boldsymbol{\theta}) := \prod_{i=1}^d L_{p_i}(\theta_i)$ and $\mathcal{P}_{\boldsymbol{\ell}} := \{ \boldsymbol{p} \in \mathbb{N}^{d}: \boldsymbol{0} \leq \boldsymbol{p} \leq \boldsymbol{P}_{\boldsymbol{\ell}}:= (\ell_1 - 1, \ell_2 - 1, \ldots, \ell_d - 1)\}$.

\subsubsection{Sensitivity-driven dimension-adaptivity}
To fully define \eqref{eq:tensor_delta_finite}, we need to specify the multiindex set, $\mathcal{L}$.
$\mathcal{L}$ is critical for the computational efficiency of constructing the sparse grid approximation: the larger the cardinality of $\mathcal{L}$ is, the larger the cost of finding the approximation.
To reduce this cost, we employ sensitivity-driven dimension-adaptivity. 

An important ingredient of dimension-adaptive algorithms \cite{GG03,He03} is a refinement indicator, $\varepsilon(\boldsymbol{\ell})$, which is used to decide which multiindices to refine in an adaptive step.
In the sensitivity-driven procedure of \cite{Fa20,Fa20b}, $\varepsilon(\boldsymbol{\ell})$ is computed in terms of unnormalized Sobol' indices \cite{So01}: we first find the equivalent spectral projection representation of the multivariate interpolation operators \eqref{eq:interp_ND}:
\begin{equation} \label{eq:spectral_to_interp}
\Ucal^{d}_{\boldsymbol{\ell}}[f^{(0)}] = \sum_{\boldsymbol{p} \in \mathcal{P}_{\boldsymbol{\ell}}} f^{(0)}(\boldsymbol{\theta}_{\boldsymbol{p}}) L_{\boldsymbol{p}}(\boldsymbol{\theta}) = 
\sum_{\boldsymbol{p} \in \mathcal{P}_{\boldsymbol{\ell}}}
c_{\boldsymbol{p}} \Phi_{\boldsymbol{p}}(\boldsymbol{\theta}),
\end{equation}
where $\Phi_{\boldsymbol{p}}(\boldsymbol{\theta}) := \prod_{i=1}^d \Phi_i(\theta_i)$ are orthonormal polynomial w.r.t.~$\pi$ and $c_{\boldsymbol{p}}$ are the spectral coefficients.
For example, if $\pi$ is the uniform distribution, $\Phi_{\boldsymbol{p}}(\boldsymbol{\theta})$ are the Legendre polynomials.
To determine the spectral coefficients $c_{\boldsymbol{p}}$, we simply solve a linear system of equations, $\sum_{\boldsymbol{p} \in \mathcal{P}_{\boldsymbol{\ell}}}
c_{\boldsymbol{p}} \Phi_{\boldsymbol{p}}(\boldsymbol{\theta}_k) = \Ucal^{d}_{\boldsymbol{\ell}}[f^{(0)}(\boldsymbol{\theta}_k)]$ for all weighted (L)-Leja points $\boldsymbol{\theta}_k$ associated to the multiindex $\boldsymbol{\ell}$ (see \cite{Fa20, Fa20b}).
Next, we rewrite the hierarchical interpolation surpluses \eqref{eq:hierarch_surplus} in terms of spectral projections:
\begin{equation*}
\Delta^{d}_{\boldsymbol{\ell}}[f^{(0)}] := \sum_{\boldsymbol{p} \in \mathcal{P}_{\boldsymbol{\ell}}} \Delta c_{\boldsymbol{p}} \Phi_{\boldsymbol{p}}(\boldsymbol{\theta}), \quad
\Delta c_{\boldsymbol{p}} := \sum_{\boldsymbol{z} \in \{0, 1\}^{d}} (-1)^{|\boldsymbol{z}|_1} c_{\boldsymbol{p} - \boldsymbol{z}}, \quad \Delta c_{\boldsymbol{0}} := c_{\boldsymbol{0}}.
\end{equation*}
It follows that
\begin{equation}\label{eq:spectral_l2_norm}
\left \| \Delta^{d}_{\boldsymbol{\ell}}[f^{(0)}] \right \|_{L^2}^2 = \sum_{\boldsymbol{p} \in \mathcal{P}_{\boldsymbol{\ell}}} \Delta c_{\boldsymbol{p}}^2 = \Delta c_{\boldsymbol{0}}^2 + \sum_{\boldsymbol{p} \in \mathcal{P}_{\boldsymbol{\ell}} \setminus \{\boldsymbol{0}\}} \Delta c_{\boldsymbol{p}}^2 = (\mathbb{E}_{\boldsymbol{\ell}}[f^{(0)}])^2 + \Delta \mathrm{Var}_{\boldsymbol{\ell}}[f^{(0)}].
\end{equation}
The first equality in \eqref{eq:spectral_l2_norm} is due to Parseval's identity, whereas the last equality follows from the orthonormality of the spectral projection basis \cite{Xi09}.
From \eqref{eq:spectral_l2_norm}, we therefore obtain the variance corresponding to each subspace.

The variance represents global information in the underlying subspace and it therefore does not provide any information about the individual parameters, their interaction, or which of these are stochastically important.
To obtain such information, we decompose $\Delta \mathrm{Var}_{\boldsymbol{\ell}}[f^{(0)}]$ further by exploiting the equivalence between spectral projection and Sobol' decompositions \cite{Su08}:
\begin{equation}\label{eq:spectral_var_decomp}
\Delta \mathrm{Var}_{\boldsymbol{\ell}}[f^{(0)}] = \sum_{i=1}^{d} \Delta \mathrm{Var}^{i}_{\boldsymbol{\ell}}[f^{(0)}] + \Delta \mathrm{Var}^{\mathrm{inter}}_{\boldsymbol{\ell}}[f^{(0)}],
\end{equation}
where
\begin{equation} \label{eq:spectral_var_indv}
\Delta \mathrm{Var}^{i}_{\boldsymbol{\ell}}[f^{(0)}] := \sum_{\boldsymbol{p} \in \mathcal{J}_{i}} \Delta c_{\boldsymbol{p}}^2, \quad \mathcal{J}_{i} := \{ \boldsymbol{p} \in \mathcal{P}_{\boldsymbol{\ell}}: \boldsymbol{p}_i \neq 0 \land \boldsymbol{p}_j = 0,  \forall j \neq i \},
\end{equation}
\begin{equation} \label{eq:spectral_var_inter}
\Delta \mathrm{Var}^{\mathrm{inter}}_{\boldsymbol{\ell}}[f^{(0)}] := \sum_{\boldsymbol{p} \in \mathcal{J}_{ \mathrm{inter}}} \Delta c_{\boldsymbol{p}}^2, \quad \mathcal{J}_{\mathrm{inter}} := \{ \boldsymbol{p} \in \mathcal{P}_{\boldsymbol{\ell}}: | \boldsymbol{p} |_0 \geq 1 \}, 
\end{equation}
where $| \boldsymbol{p} |_0$ denotes the number of non-zero entries in $\boldsymbol{p}$.
Thus, we can decompose the variance in each subspace associated to a multiindex $\boldsymbol{\ell}$ into variances associated to each individual parameters \eqref{eq:spectral_var_indv} and the variance involving all interactions between the $d$ inputs \eqref{eq:spectral_var_inter}.
Moreover, from \cite{Su08}, we know that the terms in \eqref{eq:spectral_var_decomp} are unnormalized Sobol' indices \cite{So01}, which ascertain the stochastic importance of each input and of their interaction.
Thus, the decomposition \eqref{eq:spectral_var_decomp} allows us to have detailed information about the individual inputs as well as their interaction.

We define the sensitivity-driven refinement indicator as $\varepsilon(\boldsymbol{\ell}) := s_{\boldsymbol{\ell}}$, where $s_{\boldsymbol{\ell}}$ is an integer (for each subspace) called the sensitivity index \cite{Fa20, Fa20b}.
We compute $s_{\boldsymbol{\ell}}$ as follows.
Initially, $s_{\boldsymbol{\ell}} = 0$.
Based on user-defined $d + 1$ tolerances $\boldsymbol{\tau} := (\tau_1, \tau_2, \ldots, \tau_d, \tau_{d+1})$ -- one for each individual direction ($d$ in total) and the last one for all interactions -- we compare each term in \eqref{eq:spectral_var_decomp} with the prescribed tolerances. 
If the tolerance is not exceeded, we increase  $s_{\boldsymbol{\ell}}$ by one.
Thus, $s_{\boldsymbol{\ell}}$ takes values between zero and $d + 1$.
Denote by $\boldsymbol{k}$ the multiindex associated with the subspace having the largest sensitivity score.
In each refinement step, this subspace is refined by adding all forward neighbors $\boldsymbol{k} + \boldsymbol{e}_i, \quad i = 1, 2, \ldots, d$ of $\boldsymbol{k}$, where $\boldsymbol{e}_{i}$ denotes the $i$th unit vector in $\mathbb{R}^d$, which keeps the multiindex set downward closed, i.e., without holes \cite{GG03}.
If two or more sensitivity indices are equal, we select the multiindex with the largest variance $\Delta \mathrm{Var}_{\boldsymbol{\ell}}[f^{(0)}]$.
The algorithm stops when either (i) all tolerances $\boldsymbol{\tau}$ are reached, (ii) a prescribed maximum level, $L_{\mathrm{max}}$, is reached, or (iii) there are no more multiindices to be refined.
We note that the summarized refinement procedure is inherently sequential.
In each refinement step, at most $d$ new grid points are added, i.e., at most $d$ simulations can be performed in parallel per refinement step.
Nevertheless, for functions that are smooth, the method is usually efficient, hence compensating for the restricted outer-loop parallelism.
In addition, for problems for which a significant amount of resources is needed for a single simulation, performing multiple such simulations in parallel might not be feasible.

The sensitivity-driven dimension-adaptive sparse grid algorithm was originally designed for single-fidelity uncertainty propagation settings \cite{Fa20} with the goal of estimating the expectation, variance or Sobol' indices for sensitivity analysis of outputs of interest.
Nevertheless, since the algorithm is based on interpolation, it implicitly provides a low-fidelity approximation of the high-fidelity model as well. 
Therefore, in this work we employ the sensitivity-driven algorithm to create structure-exploiting low-fidelity models in multi-fidelity settings.
Moreover, since the algorithm provides detailed information about the stochastic parameters' sensitivities, we exploit this information to additionally create reduced-dimension low-fidelity models by fixing the unimportant stochastic parameters to a deterministic value, e.g., their expectation.
We illustrate the potential of the sensitivity-driven algorithm to creating full- and reduced-dimension low-fidelity models in the following example.

\subsubsection{Illustrative example}

Consider a high-fidelity model $f^{(0)} : [0, 1]^8 \rightarrow \mathbb{R}$ that depends on eight uniformly distributed random variables $\theta_1, \theta_2, \ldots, \theta_8$:
\begin{equation} \label{eq:illustrative_case_1}
f^{(0)}(\boldsymbol{\theta}) =  1 + \cos{(\pi + \theta_1 + 0.55 \theta_2 + 0.05 \theta_3 + 0.8 \theta_4 + 0.02 \theta_5 + 0.001 \theta_6 + 0.1 \theta_7 + 0.0005 \theta_8)}.
\end{equation}

Initially, we use the sensitivity-driven algorithm to construct an eight-dimensional low-fidelity model using tolerances $\boldsymbol{\tau} = 10^{-8} \cdot \bm{1}_9$.
To exploit the algorithm to construct reduced-dimension low-fidelity models, we ascertain the sensitivity of the eight stochastic parameters provided by the 8D low-fidelity model.
The eight total Sobol' indices read 
\begin{align*}
\hat{S}_1^T = 0.5179, \quad \hat{S}_2^T = 0.1542, \quad \hat{S}_3^T = 0.0012, \quad \hat{S}_4^T = 0.3288, \\ \quad \hat{S}_5^T = 0.0002, \hat{S}_6^T = 5.8239 \times 10^{-7}, \hat{S}_6^T = 0.0050, \hat{S}_8^T = 1.4559 \times 10^{-7}.
\end{align*}
Thus, we see that $\theta_1$ is the most important parameter, $\theta_4$ the second most important and so on.
Note that the sensitivity of the eight inputs is reflected by their weights too.

Based on the information provided by the total Sobol' indices, we construct reduced models with stochastic dimension reduced from seven down to four.
We approximate the Pearson correlation coefficient between the original $f^{(0)}$ with all eight stochastic inputs \eqref{eq:illustrative_case_1} and all sensitivity-driven interpolants using $10,000$ MC samples.
The results are presented in Table \ref{tab:illustrative_ex}.
In the second column, we show which subset of inputs was used to find the low-fidelity model, in the third column, we see the estimated correlation coefficient, and in the last column, we show the number of high-fidelity model evaluations to find the sparse grid low-fidelity model.
We see that the eight-dimensional low-fidelity model is very accurate, having a correlation coefficient of $0.9999$, and is also relatively cheap to construct, requiring only $146$ evaluations of  $f^{(0)}$.
However, when the two most unimportant parameters, $\theta_6$ and $\theta_9$, are neglected, we see that the correlation coefficient virtually does not change.  
Furthermore, even if we reduce the stochastic dimension down to three, i.e., the inputs are only $\theta_1$, $\theta_2$ and $\theta_4$, the correlation coefficient is still close to one, i.e., $0.9960$, while the cost of finding the low-fidelity model decreases to only $49$ model evaluations.
\begin{table}[htbp]
\centering
\begin{tabular}{|c|c|c|c|}
\hline
	$j$ & $\boldsymbol{\theta}$ & $\rho_j$ & $\#$ high-fidelity evaluations \\
\hline
	$1$ & $\{\theta_1, \theta_2, \theta_3, \theta_4, \theta_5, \theta_6, \theta_7, \theta_8\}$ & $0.9999$ & 146 \\
\hline
$2$ & $\{\theta_1, \theta_2, \theta_3, \theta_4, \theta_5, \theta_6, \theta_7\}$ & $0.9999$ & 137 \\
\hline
$3$ & $\{\theta_1, \theta_2, \theta_3, \theta_4, \theta_5, \theta_7\}$ & $0.9999$  & 129 \\
\hline
$4$ & $\{\theta_1, \theta_2, \theta_3, \theta_4, \theta_7\}$ & $0.9998$  & 103 \\
\hline
$5$ & $\{\theta_1, \theta_2, \theta_4, \theta_7\}$ & $0.9991$  & 74 \\
\hline
$6$ & $\{\theta_1, \theta_2, \theta_4\}$ & $0.9960$ & 49 \\
\hline
\end{tabular}
\caption{Inputs to the low-fidelity models obtained by the dimension-adaptive sparse grid interpolation low-fidelity model and their properties.}
\label{tab:illustrative_ex}
\end{table}

We remark that reduced-dimension sparse grid low-fidelity models can be directly obtained from the full-dimension sparse grid low-fidelity model by fixing the unimportant inputs to a deterministic value.
However, in this case the evaluation costs of both low-fidelity models will be the same, thus restricting the potential of reduced-dimension low-fidelity models in MFMC; recall that the important ingredients in MFMC are both the correlation coefficient and the evaluation cost of low-fidelity models (see Section \ref{subsec:mfmc}).
To this end, in this work we construct reduced-dimension low-fidelity model explicitly from high-fidelity evaluations, as we did in this section.
In this way, we obtain low-fidelity model with lower evaluation times than the runtime of the full-dimension low-fidelity model.

\subsection{Low-fidelity model: Deep neural network regression}\label{subsec:ml} 
We also consider a data-driven low-fidelity model based on a feed-forward deep neural network, which is trained by solving a regression problem for a set of given training samples; see, e.g.,  \cite{Bi06,Goodfellow-et-al-2016}.

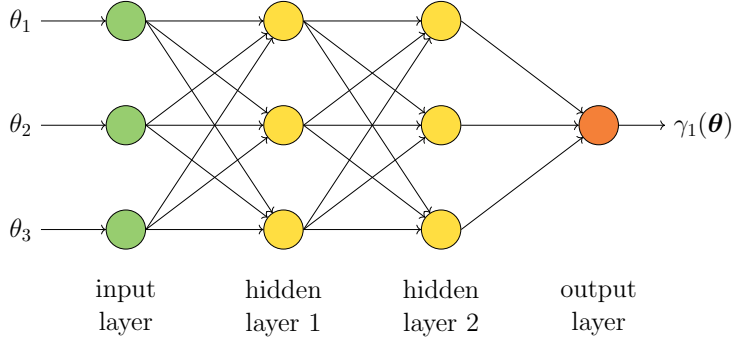
\begin{figure}[htpb]
\centering
\resizebox{0.60\columnwidth}{!}{\large\input{figures/nn}}
\caption{The structure of the neural network used as a low-fidelity model of {\sc Gene}.}
\label{fig:nn}
\end{figure}

This model is used in Section \ref{subsec:cbc_3d}, in which we consider three stochastic parameters $\theta_1, \theta_2, \theta_3$ and approximate the corresponding scalar-valued {\sc Gene} output $\gamma_1(\boldsymbol{\theta})$. 
We use a fully-connected network with two hidden layers of three neurons each and ReLU activation functions. 
This architecture is depicted in Figure~\ref{fig:nn}. 
We train the model using an RMSprop optimizer with a learning rate of $\alpha = 0.001$ that minimizes the MSE of the neural network output compared to the given training data. 
As training samples, we make use of existing data that we have available from previous numerical experiments. 
We construct and train the neural network using the Keras API on Tensorflow for \texttt{python}.

%% file: figures/nn.tex
\begin{tikzpicture}[state/.style={circle, draw, minimum size=0.75cm}]

\node[state, fill=YellowGreen] (input_1) at (0,0) {};
\node[state, fill=YellowGreen] (input_2) at (0, -2) {};
\node[state, fill=YellowGreen] (input_3) at (0, -4) {};

\node[state, fill=Goldenrod] (hidden_1_1) at (3,0) { };
\node[state, fill=Goldenrod] (hidden_1_2) at (3, -2) { };
\node[state, fill=Goldenrod] (hidden_1_3) at (3, -4) { };

\node[state, fill=Goldenrod] (hidden_2_1) at (6,0) {};
\node[state, fill=Goldenrod] (hidden_2_2) at (6, -2) {};
\node[state, fill=Goldenrod] (hidden_2_3) at (6, -4) {};

\node[state, fill=Orange] (output) at (9,-2) {};

\draw[->] (input_1.east) -- (hidden_1_1) node[at start, above right] {};
\draw[->] (input_1.east) -- (hidden_1_2) node[at start, above right] {};
\draw[->] (input_1.east) -- (hidden_1_3) node[at start, above right] {};
\draw[->] (input_2.east) -- (hidden_1_1) node[at start, above right] {};
\draw[->] (input_2.east) -- (hidden_1_2) node[at start, above right] {};
\draw[->] (input_2.east) -- (hidden_1_3) node[at start, above right] {};
\draw[->] (input_3.east) -- (hidden_1_1) node[at start, above right] {};
\draw[->] (input_3.east) -- (hidden_1_2) node[at start, above right] {};
\draw[->] (input_3.east) -- (hidden_1_3) node[at start, above right] {};

\draw[->] (hidden_1_1.east) -- (hidden_2_1) node[at start, above right] {};
\draw[->] (hidden_1_1.east) -- (hidden_2_2) node[at start, above right] {};
\draw[->] (hidden_1_1.east) -- (hidden_2_3) node[at start, above right] {};
\draw[->] (hidden_1_2.east) -- (hidden_2_1) node[at start, above right] {};
\draw[->] (hidden_1_2.east) -- (hidden_2_2) node[at start, above right] {};
\draw[->] (hidden_1_2.east) -- (hidden_2_3) node[at start, above right] {};
\draw[->] (hidden_1_3.east) -- (hidden_2_1) node[at start, above right] {};
\draw[->] (hidden_1_3.east) -- (hidden_2_2) node[at start, above right] {};
\draw[->] (hidden_1_3.east) -- (hidden_2_3) node[at start, above right] {};

\draw[->] (hidden_2_1.east) -- (output) node[at start, above right] {};
\draw[->] (hidden_2_2.east) -- (output) node[at start, above right] {};
\draw[->] (hidden_2_3.east) -- (output) node[at start, above right] {};

\node[] (param_1) at (-2, 0) {$\theta_1$};
\node[] (param_2) at (-2, -2) {$\theta_2$};
\node[] (param_3) at (-2, -4) {$\theta_3$};
\draw[->] (param_1) -- (input_1) node[at start, above right] {};
\draw[->] (param_2) -- (input_2) node[at start, above right] {};
\draw[->] (param_3) -- (input_3) node[at start, above right] {};

\node[] (result_1) at (11, -2) {$\gamma_1(\boldsymbol{\theta})$};
\draw[->] (output) -- (result_1) node[at start, above right] {};

\node[text width=1.5cm, align=center] (input_layer) at (0, -5.5) {input \\ layer};
\node[text width=1.5cm, align=center] (hidden_layer_1) at (3, -5.5) {hidden layer 1};
\node[text width=1.5cm, align=center] (hidden_layer_2) at (6, -5.5) {hidden layer 2};
\node[text width=1.5cm, align=center] (output_layer) at (9, -5.5) {output layer};
\end{tikzpicture}

%% file: sections/04_numerical_results.tex
\section{Numerical results}\label{sec:results}

In this section, we employ the proposed data-driven multi-fidelity framework for uncertainty propagation in two plasma micro-turbulence analysis scenarios to estimate the expectation and variance of the growth rate of dominant eigenmodes, $\gamma_1$ (see Section \ref{subsec:plasma}).
In Section \ref{subsec:cbc}, we consider a modified Cyclone Base Case benchmark scenario to draw our initial conclusions about the potential of our MFMC approach to quantify uncertainty in plasma micro-turbulence analysis.
First, in Section \ref{subsec:cbc_3d}, we consider the three uncertain parameters that are usually the most important for plasma micro-turbulence simulations (the log density and temperature gradients of the species).
For a more comprehensive overview of this scenario, we extend the number of uncertain inputs to eight in Section \ref{subsec:cbc_8d}.
In Section \ref{subsec:jet}, we study turbulence suppression by energetic particles with $14$ uncertain parameters.

\subsection{Modified Cyclone Base Case}\label{subsec:cbc}

We first consider a modified version of the Cyclone Base Case scenario.
In the original Cyclone Base Case, which was first presented in \cite{Di00}, only one particle species is considered, i.e., deuterium ions.
In our modified version, we have two particle species: deuterium ions and electrons \cite{Fa20,Fa20b}.
Furthermore, the magnetic geometry is described by the analytical Miller equilibrium \cite{Mi98} instead of the typically considered, simpler $\hat{s} - \alpha$ model.
We also consider collisions modelled by a linearized Landau operator.
In this way our setup more closely resembles realistic plasma micro-turbulence analysis problems.

\paragraph{Setup}
To discretize the $5$D gyrokinetic state space, we employ $184,320 := 15 \times 24 \times 1 \times 32 \times 16$ degrees of freedom in total: we use $15$ Fourier modes in the radial ($x$) direction and $24$ points along the field line, in the $z$ direction; recall that in flux-tube simulations, only one point is used in the bi-normal ($ky$) direction. 
In velocity space, we employ $32$ equidistant symmetric parallel velocity grid points and $16$ Gauss-Laguerre distributed magnetic moment points. 

The high-fidelity {\sc Gene} simulations are performed using $32$ cores on two Intel Xeon E5-2697 nodes of the CoolMUC-2 Linux cluster.\footnote{https://www.lrz.de/services/compute/linux-cluster/} 
The low-fidelity evaluations were performed on a laptop computer with an Intel Core i5-8250U CPU running at $1.60$ GHz.
All calculations were performed in double precision arithmetic.
\subsubsection{Case with three uncertain inputs}\label{subsec:cbc_3d}

First, we consider three uncertain parameters: the ion and electron logarithmic temperature gradients, $\omega_{T_i}$ and $\omega_{T_e}$, as well as their logarithmic density gradient, $\omega_n$.
Note that to ensure the quasi-neutrality assumption in plasma physics, the density gradients need to be equal.
It is known that changes in these parameters are causing the underlying micro-instability \cite{Di00} and therefore these three inputs are of interest with respect to uncertainty propagation.
We summarize the setup employed in our experiments in Table \ref{tab:param_3d}.
The parameters are modelled as uniform random variables with bounds of $\pm 25\%$ around their nominal value, which is their typical value used in typical plasma micro-turbulence numerical studies. 
\begin{table}[ht]
\centering
\begin{tabular}{|c|c|c|c|}
\hline
	 & parameter & symbol & probability distribution \\
\hline
	$\theta_1$ & ion/electron log density gradient & $\omega_n$ & $\mathcal{U}(1.6650, 2.7750)$\\
\hline
	$\theta_2$ & ion log temperature gradient & $\omega_{T_i}$ & $\mathcal{U}(7.5000, 12.5000)$\\
\hline
	$\theta_3$ & electron log temperature gradient & $\omega_{T_e}$ & $\mathcal{U}(7.5000, 12.5000)$\\
\hline
\end{tabular}
\caption{The three uncertain input parameters for the three-dimensional Cyclone Base Case and their probability distributions. By $\mathcal{U}(a,b)$ we denote a uniform distribution on $[a,b]$.}
\label{tab:param_3d}
\end{table}

For this scenario, it is known \cite{Di00} that around bi-normal wave-number $k_y \rho_s = 0.6$ there is a mode transition from an ion-temperature gradient (ITG) to a trapped electron (TEM)/electron temperature gradient (ETG) hybrid mode.
Therefore, we consider two wave numbers in our experiments: one for which the micro-instability is driven by ITG, i.e., $k_y \rho_s = 0.3$ and another one for which we have TEM/ETG micro-instability, i.e., $k_y \rho_s = 0.8$. 

\paragraph{Data-driven low-fidelity models}
For both values of $k_y \rho_s$ we construct two low-fidelity models, $f^{(1)}$ and $f^{(2)}$. 
One data-driven low-fidelity model is based on the sensitivity-driven dimension-adaptive sparse grid interpolation approximation summarized in Section \ref{subsec:sg}.
To construct it at $k_y \rho_s = 0.3$, we employ the tolerance $\bm{\tau} = (10^{-1}, 10^{-1}, 10^{-1}, 10^{-1})$.
At $k_y \rho = 0.8$, we set $\bm{\tau} = (10^{-6}, 10^{-6}, 10^{-6}, 10^{-3})$. 
We choose smaller tolerances at $k_y \rho = 0.8$ because we know from previous experiments that, compared to $k_y \rho = 0.3$, more effort is needed to obtain sufficiently accurate low-fidelity models.
Note, however, that the sparse grid model is hierarchical, therefore one can start constructing low-fidelity models with some initial tolerances and re-use those model evaluations if the tolerances need to be decreased.

In both cases, the maximum grid level is set to $L_{\mathrm{max} }= 20$. 
Only four high-fidelity evaluations with {\sc Gene} were needed to create the sparse grid low-fidelity model at $k_y \rho_s = 0.3$ and $34$ {\sc Gene} evaluations were needed for $k_y \rho_s = 0.8$.
The other low-fidelity model is based on the deep-network model summarized in Section \ref{subsec:ml}.
To train the network, we make use of existing {\sc Gene} evaluations available from previous experiments. 
At $k_y \rho_s = 0.3$, we had available $10^5$ evaluations. 
For the second wave-number, we trained on $2.5 \times 10^4$ data samples.

We show the correlation coefficients, single-core runtimes and variances of the high- and low-fidelity models in Table~\ref{tab:lo_fi_3d_0_3}.
For clarity, we use the following notation for the two low-fidelity models: the superscripts denote the fidelity of the model and to distinguish between the sparse grid and the machine learning models, we use the subscripts SG and ML, respectively.  
At $k_y \rho_s = 0.3$, the first low-fidelity model in the hierarchy is the deep-network low-fidelity model, i.e., $f^{(1)}_{\mathrm{ML}}$, and the second low-fidelity model is the sparse grid approximation, i.e., $f^{(2)}_{\mathrm{SG}}$. 
At $k_y \rho_s = 0.8$, the order of the low-fidelity models is reversed due to the values of their Pearson correlation coefficients: $f^{(1)}_{\mathrm{SG}}$ is the first low-fidelity model and $f^{(2)}_{\mathrm{ML}}$ is the second low-fidelity model.
\begin{table}[ht]
    \centering
    \begin{tabular}{cc}
    \begin{tabular}{|c|c|c|c|}
        \hline
        $f^{(j)}$ & $\rho_j$ & $w_j~[sec]$ & $\sigma_j^2$\bigstrut[t] \\
        \hline
        $f^{(0)}$ & 1.0000 & 260.1697 & 0.0170\bigstrut[t]  \\
        \hline
        $f^{(1)}_{\mathrm{ML}}$ & 0.9998 & 0.0019 & 0.0165 \bigstrut[t]\\
        \hline
        $f^{(2)}_{\mathrm{SG}}$ & 0.9989 & 0.0008 & 0.0144 \bigstrut[t]\\
        \hline
    \end{tabular} &
    \begin{tabular}{|c|c|c|c|}
        \hline
        $f^{(j)}$ & $\rho_j$ & $w_j~[sec]$ & $\sigma_j^2$\bigstrut[t] \\
        \hline
        $f^{(0)}$ & 1.0000 & 240.5123 & 0.0754 \bigstrut[t] \\
        \hline
        $f^{(1)}_{\mathrm{SG}}$ & 0.9819 & 0.0166 & 0.0747\bigstrut[t] \\
        \hline
        $f^{(2)}_{\mathrm{ML}}$ & 0.9708 & 0.0017 & 0.0703\bigstrut[t] \\
        \hline
    \end{tabular}\\
    ~ & ~\\
    (a) $k_y \rho_s = 0.3$ & (b) $k_y \rho_s = 0.8$
    \end{tabular}
    \caption{Correlation and costs of high-fidelity and low-fidelity models for  modified Cyclone Base Case.}
    \label{tab:lo_fi_3d_0_3}
\end{table}
The three quantities were estimated using $1,000$ evaluations of the high- and low-fidelity models.
The evaluation costs of the low-fidelity models are $4-5$ orders of magnitude smaller compared to the costs of the high-fidelity model. 
At $k_y \rho_s = 0.3$, both low-fidelity models have correlation coefficients close to $1.0$, whereas at $k_y \rho_s = 0.8$, the low-fidelity models are less accurate and thus poorer correlated to the high-fidelity model.

\paragraph{Estimating the expectation}
We employ MFMC to estimate the expectation and variance of the growth rate at both $k_y \rho_s = 0.3$ and $k_y \rho_s = 0.8$ for budgets $p \in \{5 \times 10^2, 10^3, 5 \times 10^3, 10^4, 5 \times 10^4 \}$ seconds.
For comparison purposes, we also compute MC approximations for the same budgets $p$.

We first compare the MC and MFMC expectation estimators in terms of their MSE that is estimated from  $N$ replicates as:
\begin{equation}\label{eq:mse_replicates}
    e_{\text{MSE}}(\hat{E}^{(\cdot)}_n) = \frac{1}{N} \sum_{n=1}^{N} (\hat{\mu}_{\text{ref}} - \hat{E}^{(\cdot)}_n)^2,
\end{equation}
where $\hat{\mu}_{\text{ref}}$ serves as the reference mean estimator and $\hat{E}^{(\cdot)}_n$ is either an MFMC or an MC estimator.
The reference $\hat{\mu}_{\text{ref}}$ is obtained with MFMC using a large budget $p_{\text{ref}} = 10^5$ seconds. 
At $k_y \rho_s = 0.3$, the reference is $\hat{\mu}_{\text{ref}} = 0.6811$ and at $k_y \rho_s = 0.8$, we obtain $\hat{\mu}_{\text{ref}} = 0.5560$.

\begin{figure}[ht]
\centerfloat
\begin{minipage}[ht]{0.5\textwidth}
\includegraphics[width=\textwidth]{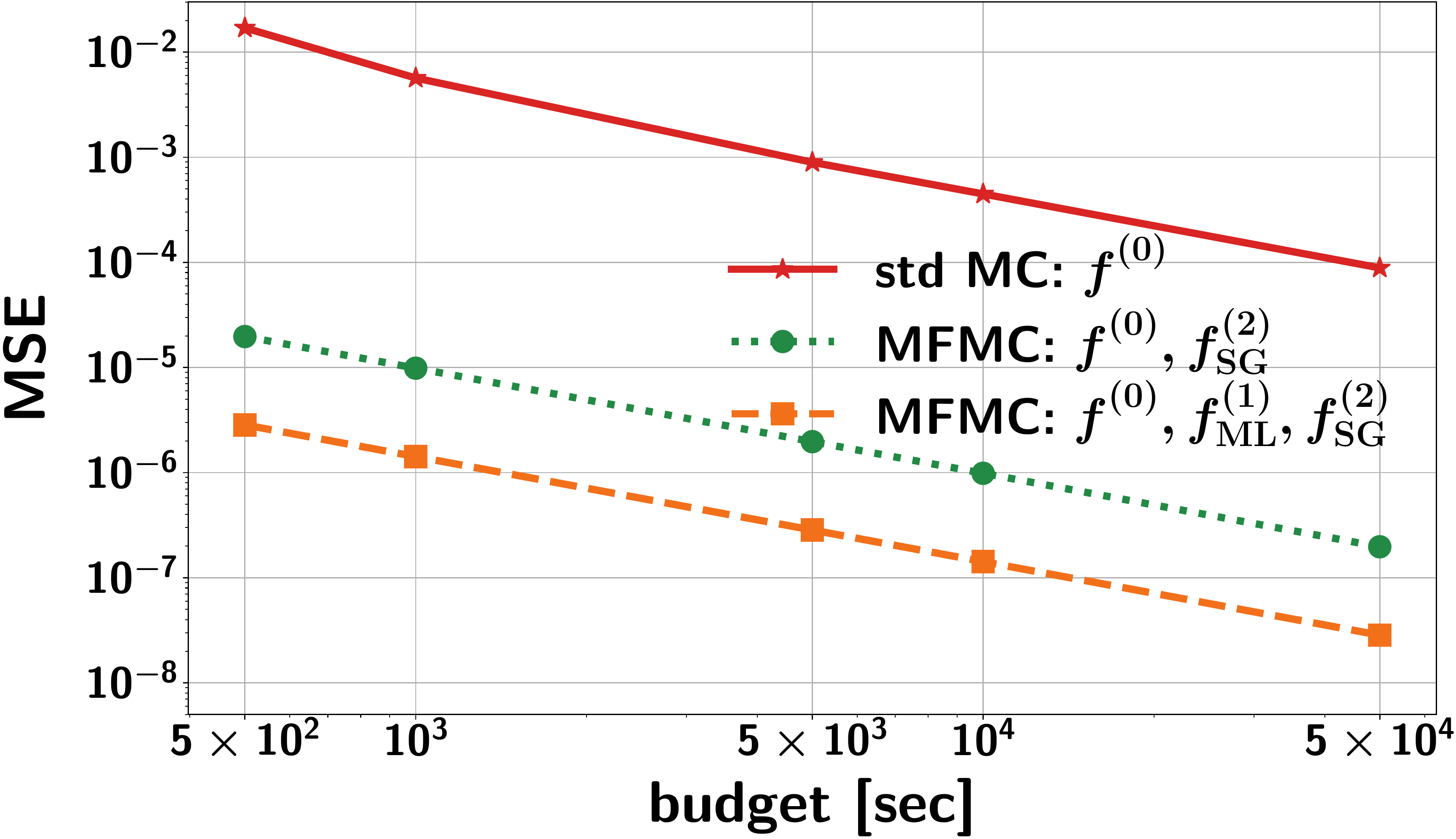}
\end{minipage}
\begin{minipage}[ht]{0.5\textwidth}
\includegraphics[width=\textwidth]{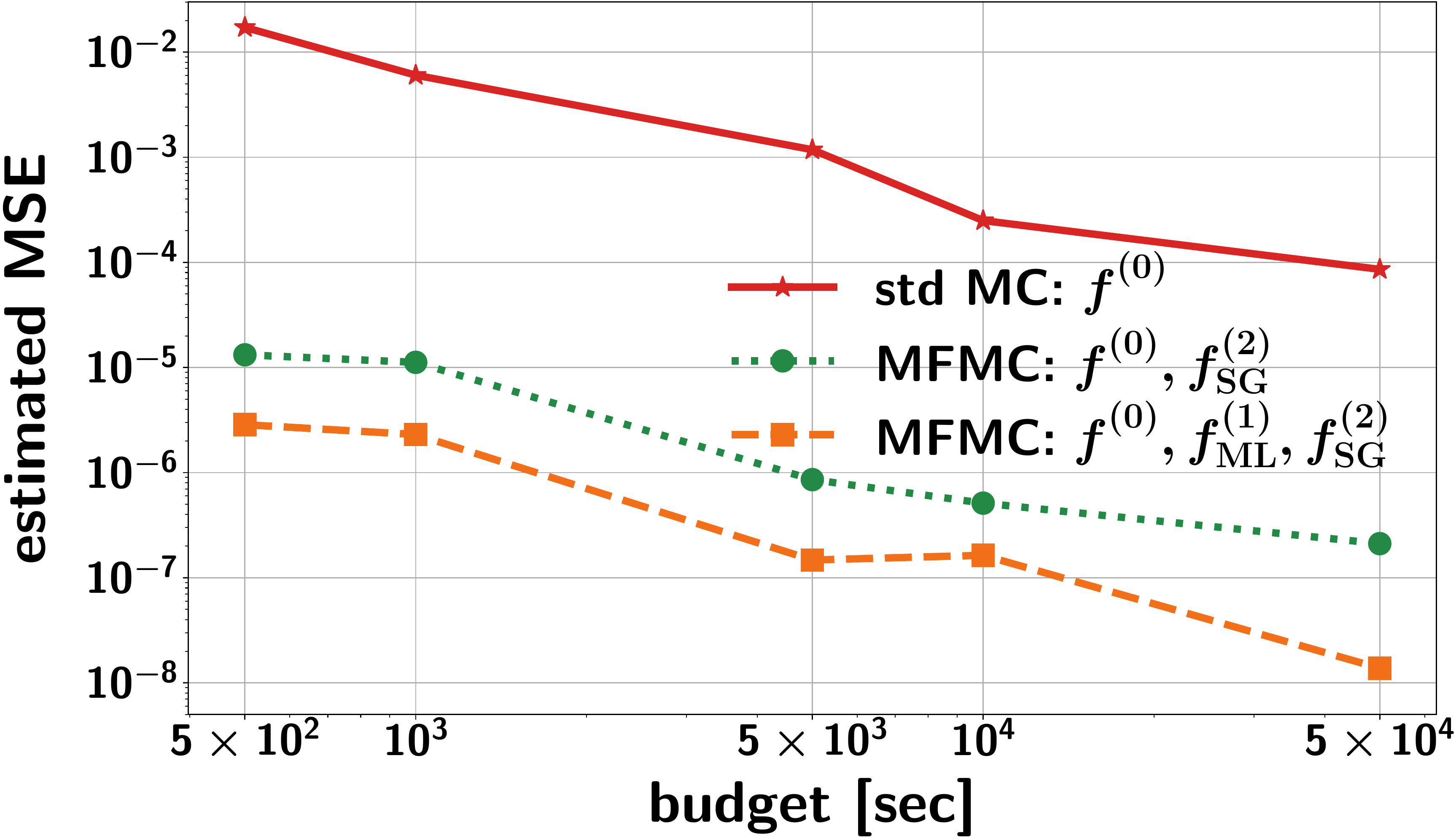}
\end{minipage}
\caption{Analytical MSE (left) and estimated MSE (right) of standard MC and MFMC for the Cyclone Base Case with 3 uncertain inputs and $k_y \rho_s = 0.3$.}
\label{fig:CBC_3D_0_3}
\end{figure}
\begin{figure}[ht]
\centerfloat
\begin{minipage}[ht]{0.5\textwidth}
\includegraphics[width=\textwidth]{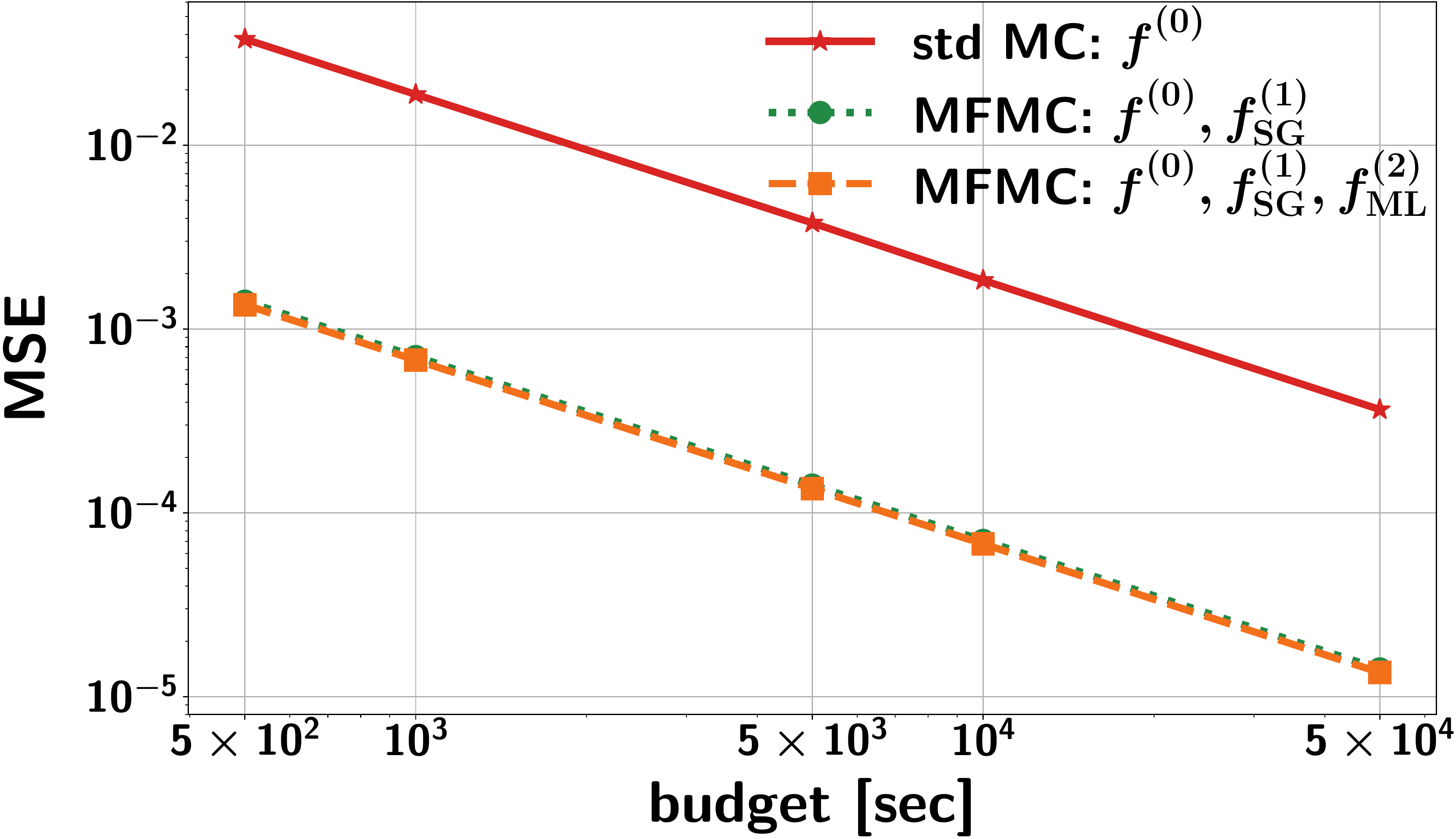}
\end{minipage}
\begin{minipage}[ht]{0.5\textwidth}
\includegraphics[width=\textwidth]{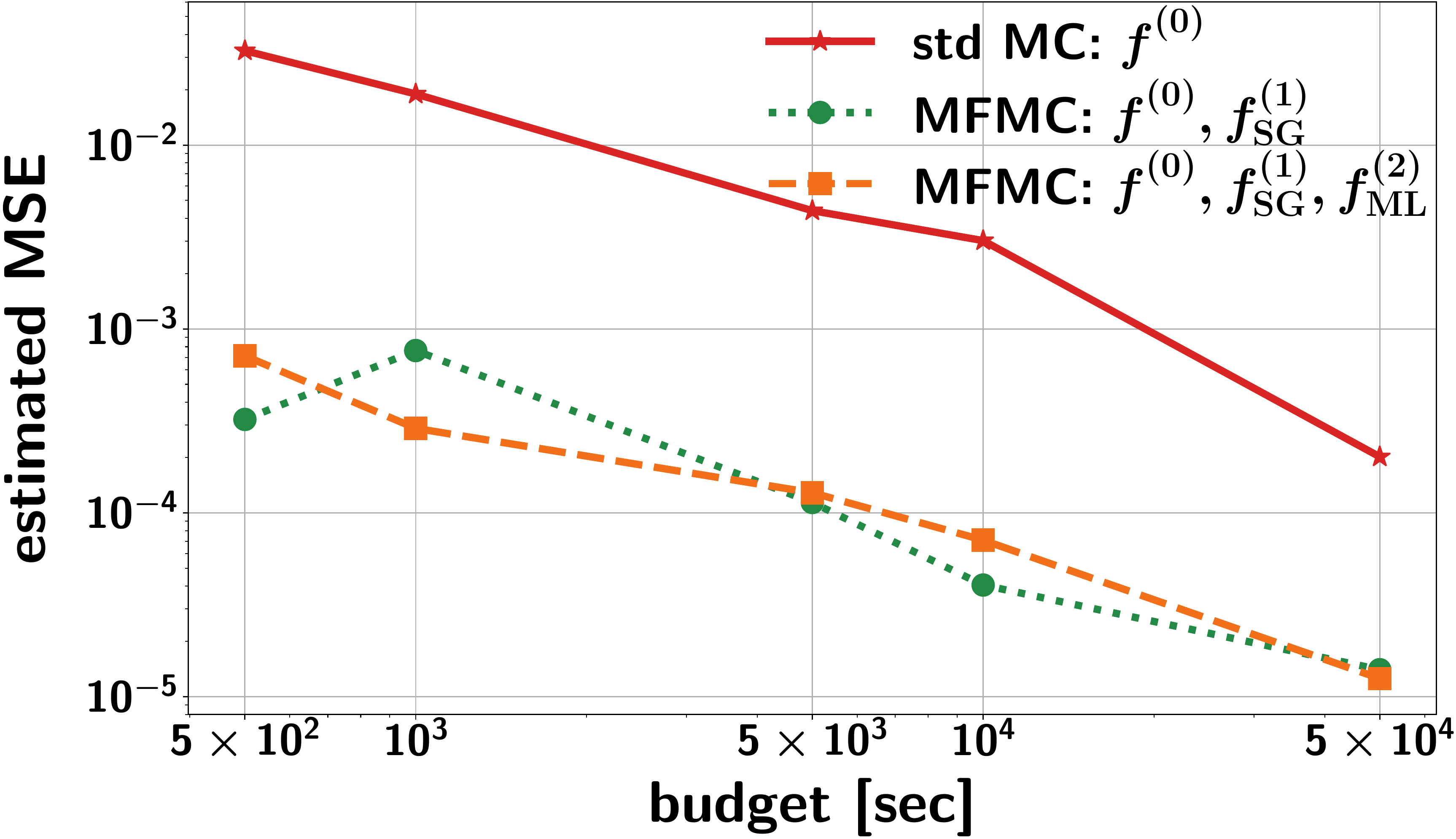}
\end{minipage}
\caption{Analytical MSE (left) and estimated MSE (right) of standard MC and MFMC for the Cyclone Base Case with 3 uncertain inputs and $k_y \rho_s = 0.8$.}
\label{fig:CBC_3D_0_8}
\end{figure}
For a broader overview, we consider two MFMC estimators: one in which we have the sparse grid low-fidelity model only and another in which we consider both the sparse grid and the machine-learning-based low-fidelity models.
We first compute the analytic MSE obtained via \eqref{eq:mse_mc} and \eqref{eq:mse_mfmc} based on the analysis in \cite{PWG16}. The analytic MSE is obtained from the measured values in Table~\ref{tab:lo_fi_3d_0_3} and gives an indication what variance reduction and speedup we expect to see in the numerical simulations. Notice that the analytical MSEs are computed purely from the values given in Table~\ref{tab:lo_fi_3d_0_3} and require no additional numerical simulations. Additional analyses of the MFMC approach to estimate how much variance reduction one can expected before conducing numerical experiments are presented in \cite{PWK16MFMCAsymptotics}. The analytic MSEs are shown in the left panel of Figure~\ref{fig:CBC_3D_0_3} and~\ref{fig:CBC_3D_0_8}.
The estimated MSE, computed via \eqref{eq:mse_replicates} using $N = 10$ replicates, are visualized in the right panel in each figure.
MFMC is around three orders of magnitude more accurate in terms of the estimated MSE for $k_y \rho_s = 0.3$ and about one order of magnitude more accurate for $k_y \rho_s = 0.8$, compared to the MC estimator.
The results in Figure \ref{fig:CBC_3D_0_3} for $k_y \rho_s = 0.3$ indicate that adding the machine learning model to MFMC decreases the MSE further.
This is explained by the high correlation and the low evaluation costs of the two low-fidelity models as shown in Table \ref{tab:lo_fi_3d_0_3}.
In contrast, for $k_y \rho_s = 0.8$, the results in Figure \ref{fig:CBC_3D_0_8} indicate that adding the machine learning model leads to little improvement, which is explained by the poorer correlation coefficient as shown in Table \ref{tab:lo_fi_3d_0_3}.

\begin{figure}[ht]
\centerfloat
\begin{minipage}[ht]{0.5\textwidth}
\includegraphics[width=\textwidth]{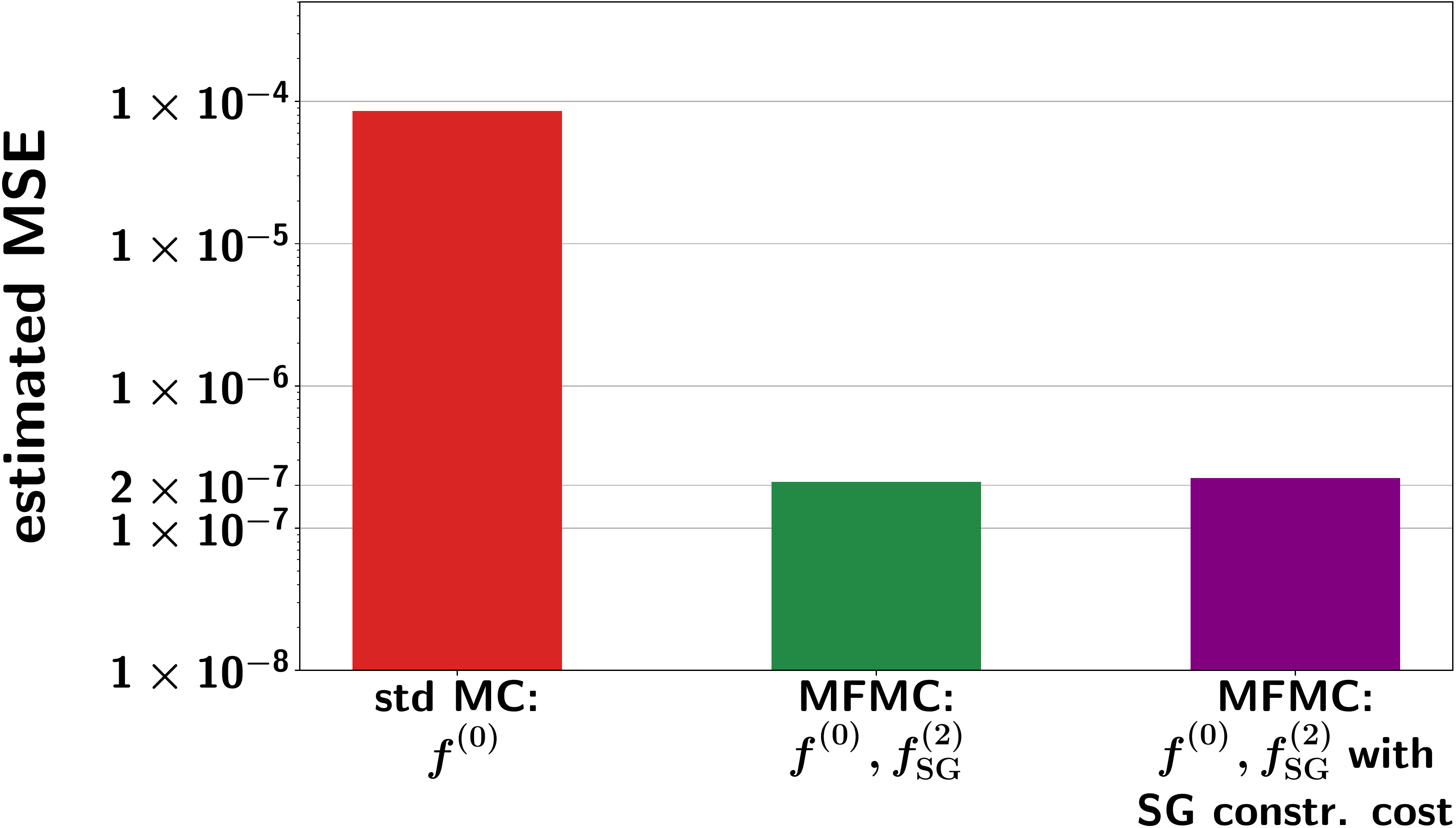}
\end{minipage}
\begin{minipage}[ht]{0.5\textwidth}
\includegraphics[width=\textwidth]{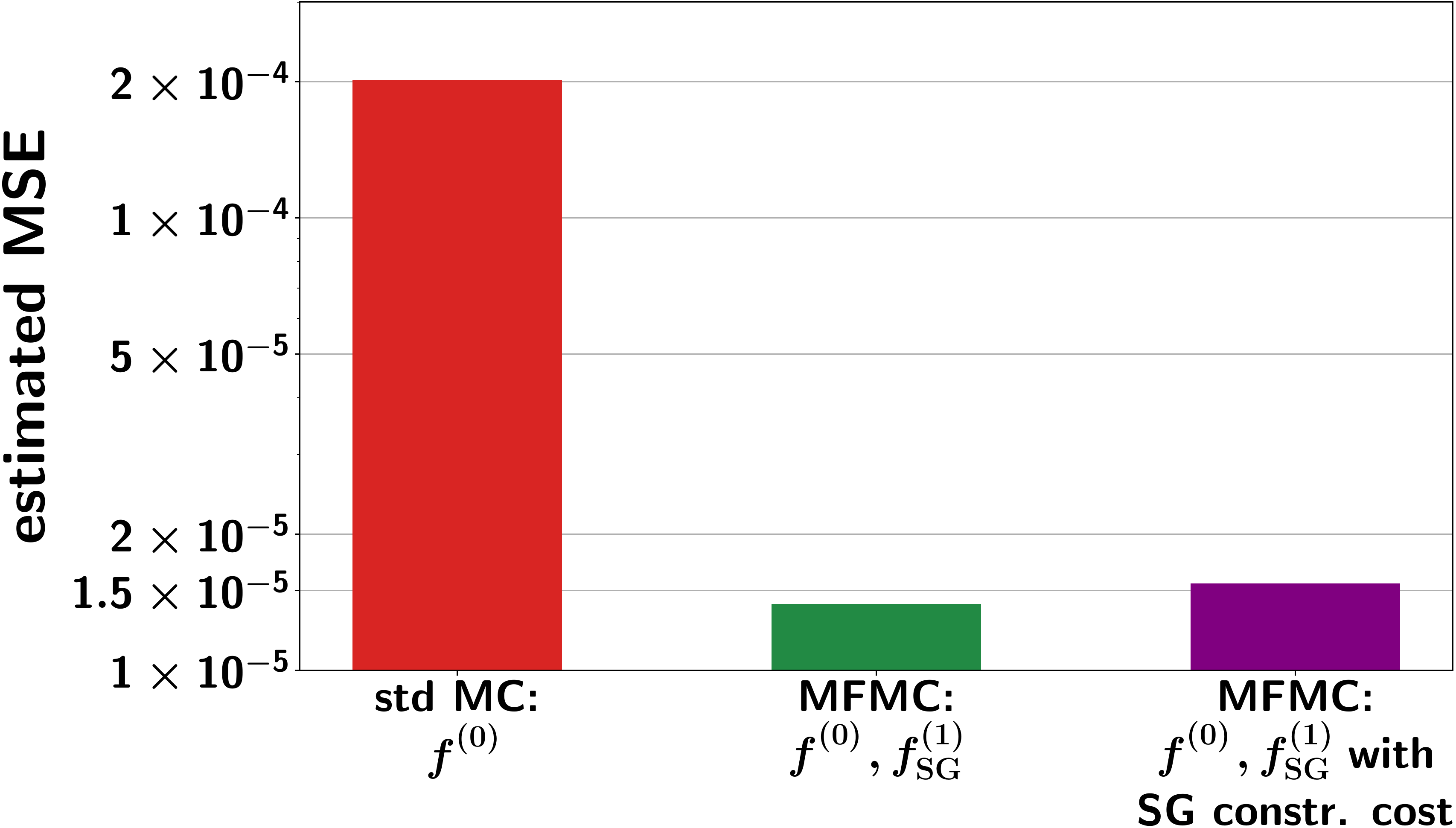}
\end{minipage}
\caption{MSE of mean estimators for $p=5 \times 10^4$ for standard MC and MFMC using the low-fidelity model $f^{(1)}_{\text{SG}}$ with and without considering the construction cost of the sparse grid model for the Cyclone Base Case with three uncertain inputs and $k_y \rho_s = 0.3$ (left) and $k_y \rho_s = 0.8$ (right).}
\label{fig:CBC_3D_0_8_with_offline_cost}
\end{figure}

We also compute the MSE of the MFMC estimator in which we explicitly account for the construction cost of low-fidelity models.
We compute the estimator which depends on the high-fidelity and the sparse grid low-fidelity model and focus on the estimator with budget $p = 5 \times 10^4$.
At $k_y \rho_s = 0.3$, four high-fidelity evaluations, i.e., $4 \times w_0 = 4 \times 260.1697 \approx 1,040$ seconds were required to construct the sparse grid low-fidelity model $f^{(2)}_{\mathrm{SG}}$.
We subtract the construction budget from the total budget of $5 \times 10^4$ seconds and use the remaining budget $p' = 5 \times 10^4 - 1,040 = 48,960$ seconds for MFMC sampling, i.e., around $98 \%$ of the original budget.
Analogously, at $k_y \rho_s = 0.8$ the budget used to construct the low-fidelity model $f^{(1)}_{\mathrm{SG}}$ was $34 \times w_0 \approx 5,772$ seconds.
The remaining budget for MFMC sampling is therefore $p' = 5 \times 10^4 - 5,772 = 44,228$ seconds, i.e., roughly $88 \% $ of the original $p$.
We visualize the corresponding results in Figure~\ref{fig:CBC_3D_0_8_with_offline_cost} (left, $k_y \rho_s = 0.3$; right, $k_y \rho_s = 0.8$).
We compare the MSE of standard MC sampling, standard MFMC sampling (without offline cost) and MFMC sampling in which we explicitly consider the construction cost. 
As expected, subtracting the sparse grid construction cost of only four high-fidelity evaluations at $k_y \rho_s = 0.3$ has a negligible effect on the MSE of the MFMC estimator.
At $k_y \rho_s = 0.8$, subtracting the construction cost of the low-fidelity model has an insignificant effect as well: the MFMC estimator remains more than one order of magnitude more accurate than the standard MC estimator despite the fact that the sampling budget was decreased by roughly $12 \%$. 

\paragraph{Speedup} To put the computational savings provided by our MFMC approach into perspective, let us take a closer look at how the obtained variance reduction translates into computational speedups.
High-fidelity simulations are performed on 32 cores, whereas low-fidelity simulations are performed on a single core only because they are so much cheaper. 
At both $k_y \rho_s = 0.3$ and $k_y \rho_s = 0.8$, the MFMC estimator with the smallest MSE was obtained for budget $p = 5 \times 10^4$ seconds, hence this will serve as reference in the following. 
Notice that the budget of $p = 5 \times 10^4$ seconds on a single core translates into $ 5 \times 10^4 / 60 / 32 \approx 26$ minutes of high- and low-fidelity simulations on $32$ cores.
Computing a standard MC estimator with the same MSE as the MFMC estimator requires a budget of about $104,583,190$ seconds on a single core, which corresponds to $104,583,190 / 3,600 / 24 / 32 \approx 38$ days of high-fidelity simulations on 32 cores.
Finally, the MFMC estimator with only the sparse grid low-fidelity model needs a budget of about $249,916$ seconds, i.e., about $130$ minutes of high- and low-fidelity simulations in 32 cores.
To summarize, at $k_y \rho_s = 0.3$, the variance reduction obtained by our MFMC approach leads to a runtime reduction from $38$ days to $26$ minutes on 32 cores if both low-fidelity models are used.
At $k_y \rho_s = 0.8$, a budget of about $1,339,482$ seconds on a single core is needed to obtain a standard MC estimator with the same MSE as the MFMC estimator with budget $p = 5 \times 10^{4}$ on a single core, which means that a standard MC estimator would require about $12$ hours of high-fidelity simulations on 32 cores.
In contrast, MFMC requires only $26$ minutes of high- and low-fidelity simulations on 32 cores.
Hence, we see that even though the low-fidelity models at $k_y \rho_s = 0.8$ were not as accurate as at $k_y \rho_s = 0.3$, using them in the context of MFMC lead to a runtime reduction from $12$ hours to $26$ minutes on 32 cores.

\begin{figure}[ht]
\centerfloat
\begin{minipage}[ht]{0.5\textwidth}
\includegraphics[width=\textwidth]{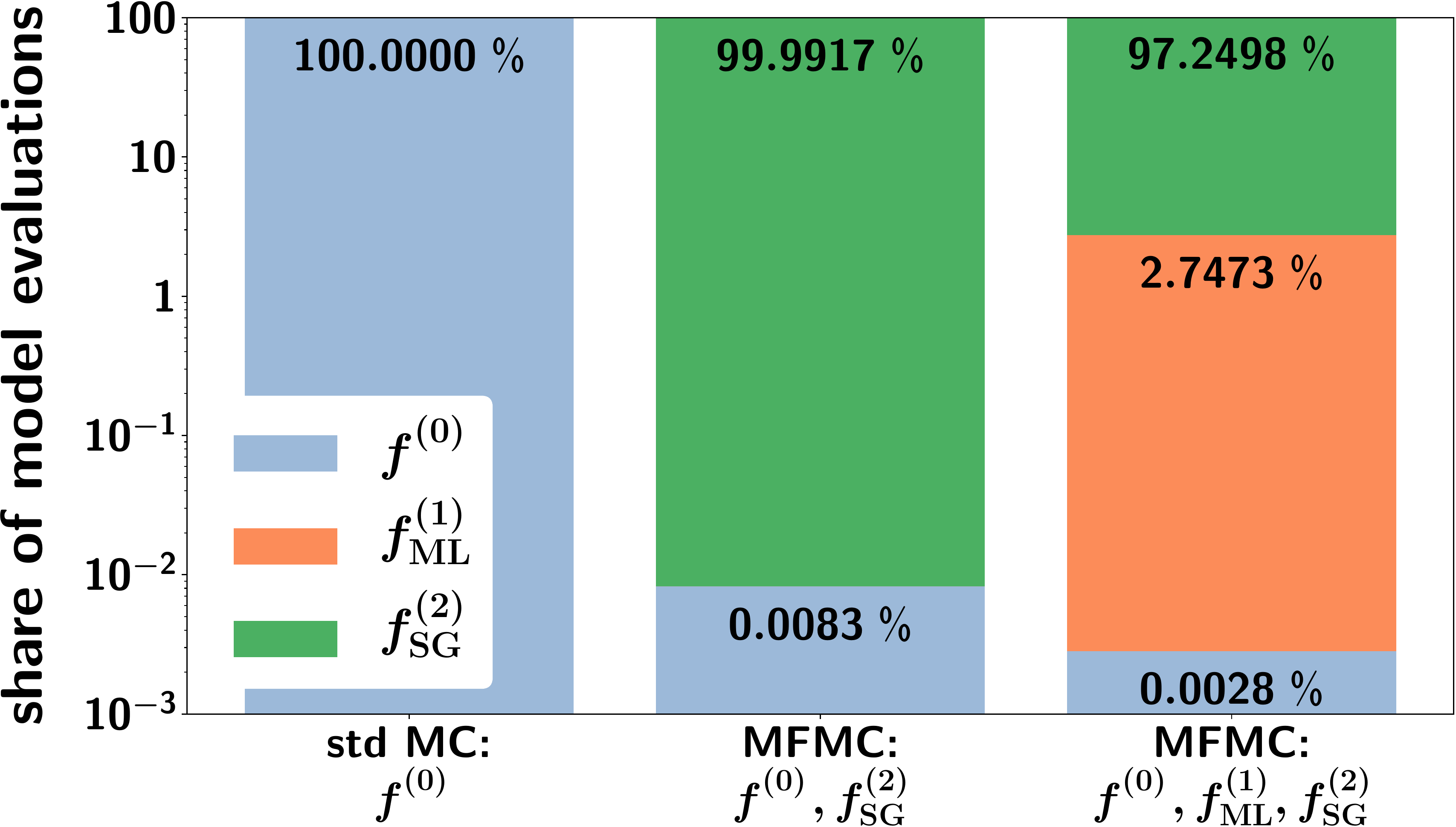}
\end{minipage}
\begin{minipage}[ht]{0.5\textwidth}
\includegraphics[width=\textwidth]{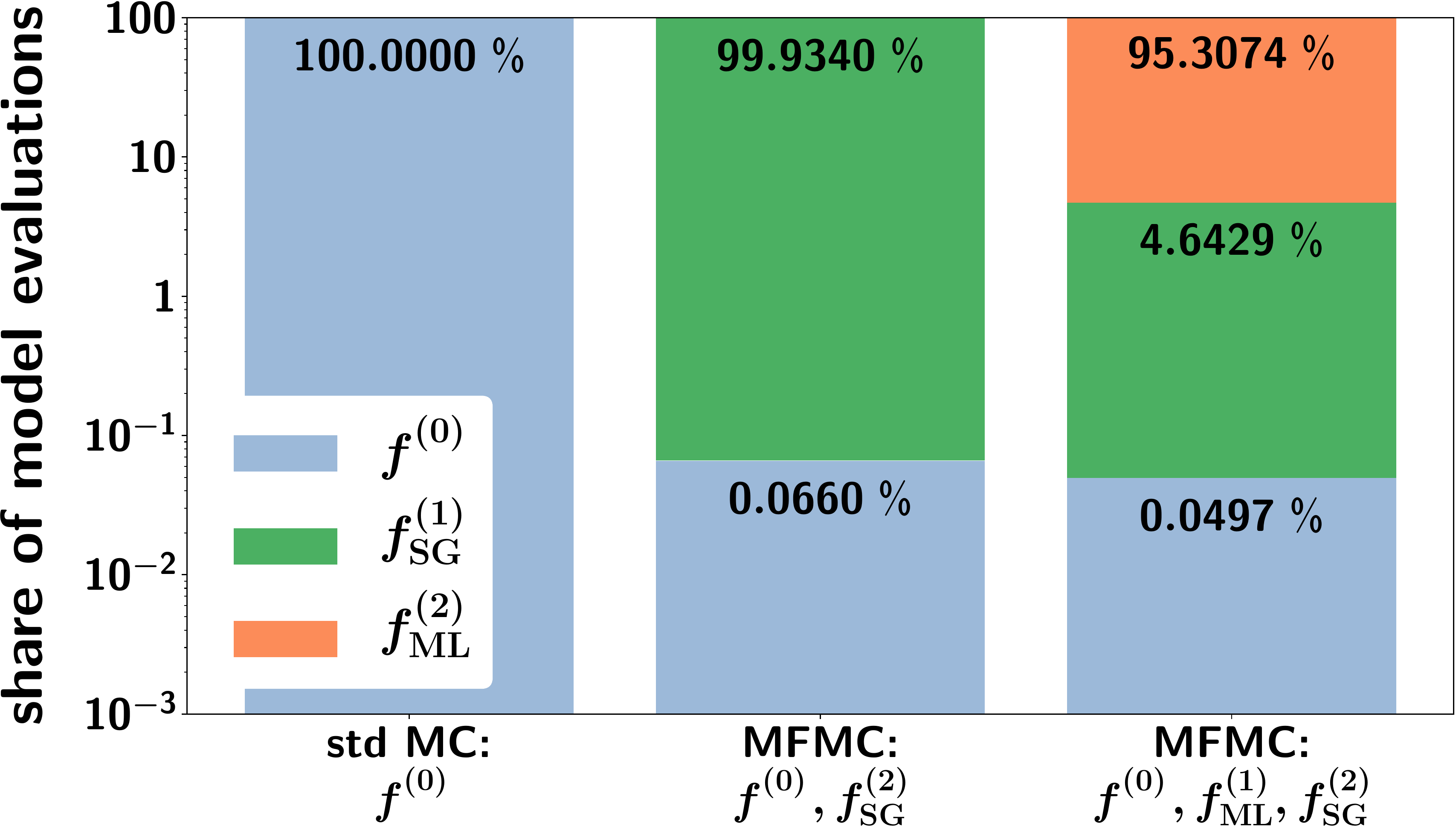}
\end{minipage}
\caption{The distribution of evaluations over the models for standard MC and MFMC with two and three models for $k_y \rho_s = 0.3$ (left) and $k_y \rho_s = 0.8$ (right). On the logarithmic y-axis, the percentage of evaluations of the different models is given.}
\label{fig:evals_3D}
\end{figure}

\paragraph{Observations} The reported results demonstrate two aspects of MFMC: first, even models with limited accuracy can be useful for MFMC and can lead to orders of magnitude improvements in terms of MSE compared to standard MC estimators.
Especially in the context of plasma physics, where one high-fidelity simulation can require significant computational resources, this MFMC feature can make the estimation of the expectation (or of other quantities of interest, such as variance or Sobol' indices for sensitivity analysis) computationally feasible.
Second, adding more than one low-fidelity model to MFMC does not necessarily lead to a further reduction of the MSE but depends on the properties of the low-fidelity model.
Figure~\ref{fig:evals_3D} shows the distribution of the number of samples among the models. 
At most $0.0497 \%$ of the total number of samples are with the high-fidelity model in MFMC, whereas standard MC sampling employs the high-fidelity model exclusively.
This provides a significant computational advantage compared to single-fidelity sampling MC schemes.

\paragraph{Estimating the variance} Table~\ref{tab:mean_var_3d_0_3} shows the variance estimates obtained in our experiments. We used the same low-fidelity models as for the expectation estimation.
To distinguish between the MFMC estimators using one and two low-fidelity models, we use the superscripts $MFMC, 1$ and $MFMC, 2$, respectively. 
For $k_y \rho = 0.3$, no estimate is obtained with budget $p = 500$ because at least two samples are needed to estimate the variance and one evaluation of the high-fidelity model takes already $260.1698 > p/2$ seconds; cf.~Table \ref{tab:lo_fi_3d_0_3}.
Using the same setup as for estimating the reference means, we obtain the reference variance $\hat{\sigma}^2_{ref} = 0.016765$ at $k_y \rho_s = 0.3$ and $\hat{\sigma}^2_{\text{ref}} = 0.0757$ at $k_y \rho_s = 0.8$.  
We observe that again MFMC yields more accurate results than the standard MC estimator: at $k_y \rho_s = 0.3$, both MFMC estimators have an extra digit of accuracy, whereas at $k_y \rho_s = 0.8$, the MC and MFMC estimators have two digits of accuracy for most budgets, but the MFMC estimator with one low-fidelity model yields a variance similar to the reference.
\begin{table}[ht]
\centering    
\begin{tabular}{cc}
\begin{tabular}{|c|c|c|c|}
\hline
	$p$ & $\widehat{\Var}^{MC}$ & $\widehat{\Var}^{MFMC, 1}$ & $\widehat{\Var}^{MFMC, 2}$\bigstrut[t] \\
\hline
	500 & - & - & -\\
\hline
	1,000 & 0.015314 & 0.017189 & 0.016677 \\
\hline
	5,000 & 0.015497 & 0.016826 & 0.016725 \\
\hline
	10,000 & 0.017299 & 0.016873 & 0.016762 \\
\hline
	50,000 & 0.017132 & 0.016714 & 0.016769 \\
\hline
\end{tabular} & \begin{tabular}{|c|c|c|c|}
\hline
	$p$ & $\widehat{\Var}^{MC}$ & $\widehat{\Var}^{MFMC, 1}$ & $\widehat{\Var}^{MFMC, 2}$ \bigstrut[t]\\
\hline
	500 & 0.0749 & 0.0753 & 0.0724 \\
\hline
	1,000 & 0.0912 & 0.0723 & 0.0805 \\
\hline
	5,000 & 0.0695 & 0.0751 & 0.0801 \\
\hline
	10,000 & 0.0702 & 0.0738 & 0.0727 \\
\hline
	50,000 & 0.0759 & 0.0757 & 0.0750 \\
\hline
\end{tabular}\\
~ & ~\\
(a) $k_y \rho_s = 0.3$, reference $\hat{\sigma}^2_{ref} = 0.016765$  & (b) $k_y \rho_s = 0.8$, reference $\hat{\sigma}^2_{\text{ref}} = 0.0757$
\end{tabular}
\caption{The estimated variance of the output of interest at $k_y \rho_s = 0.3$ (left) and $k_y \rho_s = 0.8$ (right) obtained with MC and MFMC estimators in the Cyclone Base Case with 3 uncertain inputs. The superscripts $MFMC, 1$ and $MFMC, 2$ refer to estimators obtained from MFMC using one and two low-fidelity models, respectively.}
\label{tab:mean_var_3d_0_3}
\end{table}

\subsubsection{Cyclone Base Case with eight uncertain inputs}\label{subsec:cbc_8d}

For a broader overview of this scenario, we extend the setup from three to eight uncertain parameters.
Besides the logarithmic temperature and density gradients, we consider five additional uncertain parameters: the ion-electron temperature ratio, $T_i / T_e$; $\beta$, the ratio of kinetic to magnetic pressure inside the plasma; the normalized collision frequency, $\nu_c$; the safety factor, $q$, which describes the relationship of the number of toroidal turns of a magnetic field line to number of poloidal turns; and its derivative, the magnetic shear, $\hat{s}$.
We model these eight inputs as independent uniform variables with bounds of $25\%$ around their nominal values, as shown in Table \ref{tab:param_8d}.
\begin{table}[ht]
\centering
\begin{tabular}{|c|c|c|c|}
\hline
	& parameter & symbol & probability distribution \\
\hline
$\theta_1$ & ion/electron log density gradient & $\omega_n$ & $\mathcal{U}(1.665, 2.775)$\\
\hline
	$\theta_2$ & ion log temperature gradient & $\omega_{T_i}$ & $\mathcal{U}(7.500, 12.500)$\\
\hline
	$\theta_3$ & electron log temperature gradient & $\omega_{T_e}$ & $\mathcal{U}(7.500, 12.500)$\\
	\hline
	$\theta_4$ & plasma beta & $\beta$ & $\mathcal{U}(0.598 \times 10^{-3}, 0.731 \times 10^{-3})$ \bigstrut[t]\\
\hline
	$\theta_5$ & collision frequency & $\nu_c$ & $\mathcal{U}(0.238 \times 10^{-2}, 0.322 \times 10^{-2})$ \bigstrut[t]\\
\hline
	$\theta_6$ & temperature ratio & $T_i / T_e$ & $\mathcal{U}(0.950, 1.050)$\\
\hline
	$\theta_7$ & magnetic shear & $\hat{s}$ & $\mathcal{U}(0.716, 0.875)$ \\
\hline
	$\theta_8$ & safety factor & $q$ & $\mathcal{U}(1.330, 1470)$ \\
\hline
\end{tabular}
\caption{The eight uncertain input parameters for the eight-dimensional Cyclone Base Case and their probability distributions.}
\label{tab:param_8d}
\end{table}

We focus on one perpendicular wave number, i.e., $k_y \rho_s = 0.3$.
In MFMC, we consider only the sensitivity-driven sparse grid low-fidelity model summarized in Section \ref{subsec:sg}.
We do not employ the deep-network-based low-fidelity model because we do not have available training data from previous experiments. 
Running numerical simulations to generate sufficient training data is beyond our means for this work.

\paragraph{Data-driven low-fidelity model}
To find the low-fidelity sparse grid model $f^{(1)}$, we prescribe tolerances $\bm{\tau} = 10^{-6} \cdot \bm{1}_9$ and a maximum level of refinement $L_{\mathrm{max}} = 20$. 
With this setup, we need only $52$ {\sc Gene} evaluations to construct the low-fidelity model.
Table~\ref{tab:lo_fi_8d_0_3} reports the correlation coefficients, single-core runtimes and variances of the high- and low-fidelity models, obtained using $1,000$ MC evaluations from both models.
\begin{table}[ht]
    \centering
    \begin{tabular}{|c|c|c|c|}
        \hline
        $f^{(j)}$ & $\rho_j$ & $w_j~[sec]$ & $\sigma_j^2$ \bigstrut[t]\\
        \hline
        $f^{(0)}$ & 1.0000 & 257.0304 & 0.0168\bigstrut[t] \\
        \hline
        $f^{(1)}$ & 0.9999 & 0.1006 &  0.0169\bigstrut[t] \\
        \hline
    \end{tabular}
    \caption{Overview of the high-fidelity and low-fidelity models for $k_y \rho_s = 0.3$ used in the MFMC estimation in the eight-dimensional Cyclone Base Case.}
    \label{tab:lo_fi_8d_0_3}
\end{table}
We see that the low-fidelity model is highly correlated with a correlation coefficient of $0.9999$, while achieving four orders of magnitude cost reduction.

\paragraph{Estimating the expectation}
We consider budgets $p \in \{5 \times 10^3, 10^4, 5 \times 10^4\}$ seconds.
Compared to the case with three inputs, we increased the smallest budget from $500$ to $5,000$ seconds to allow for at least one high-fidelity evaluation in the MFMC estimator.
We compare the standard MC and MFMC expectation estimators with respect to the reference $\hat{\mu}_{\text{ref}} = 0.6794$, computed using MFMC with a budget $p_{\text{ref}} = 10^5$ seconds.

\begin{figure}[ht]
\centerfloat
\begin{minipage}[ht]{0.5\textwidth}
\includegraphics[width=\textwidth]{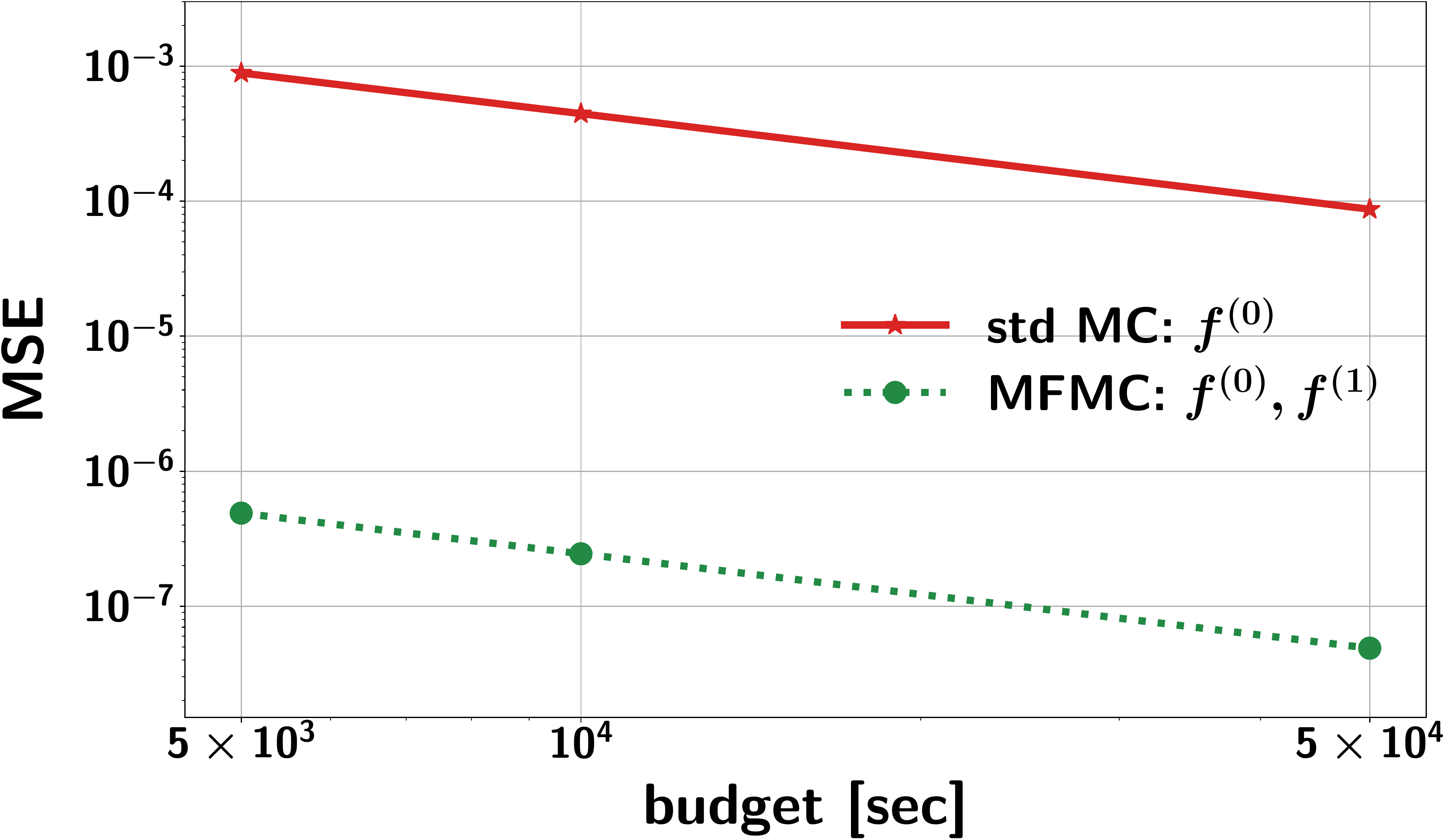}
\end{minipage}
\begin{minipage}[ht]{0.5\textwidth}
\includegraphics[width=\textwidth]{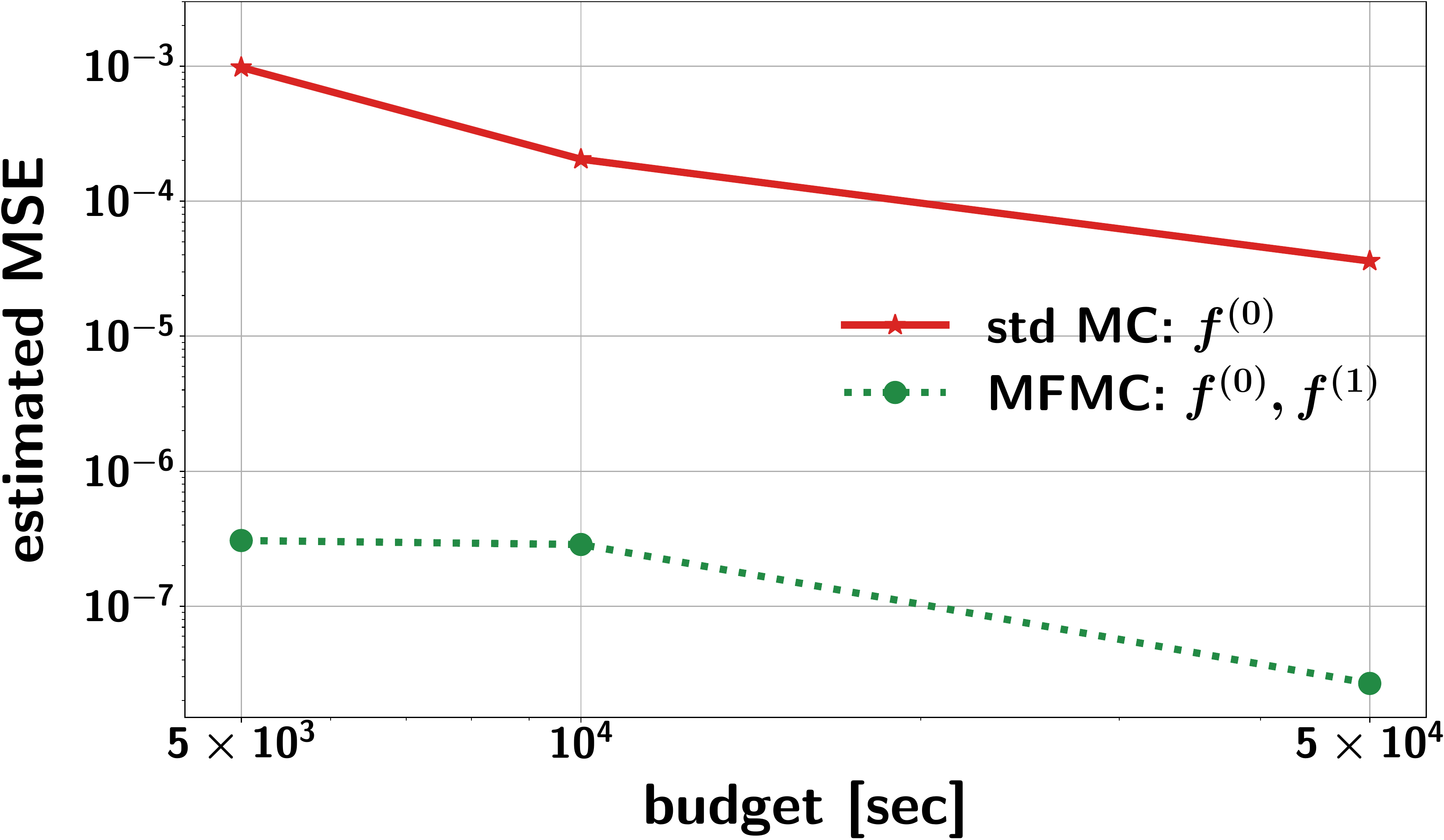}
\end{minipage}
\caption{Analytical MSE of the mean estimator (left) and estimated MSE of standard MC and MFMC results (right) for the 8-dimensional Cyclone Base Case for $k_y \rho_s = 0.3$.}
\label{fig:CBC_8D_0_3}
\end{figure}

In Figure \ref{fig:CBC_8D_0_3}, we compare MC and MFMC in terms of both the analytic (left) and the estimated MSE \eqref{eq:mse_replicates} computed using $N = 10$ replicates (right).
The MFMC estimate is about three orders of magnitude more accurate than the standard MC estimate for the same budget.
This speedup is achieved because an accurate low-fidelity model is used that is much cheaper to evaluate than the high-fidelity model.
\paragraph{Speedup}
This translates into the following runtime reduction: to obtain an MSE similar to the smallest MSE, i.e., the MSE of our MFMC estimator with budget $5 \times 10^4$ seconds on a single core, the standard MC estimator requires a budget of about $43,447,972$ seconds on a single core.
Using $32$ cores per high-fidelity simulation, this budget requires about $16$ days of simulations.
In contrast, our MFMC approach requires only about $26$ minutes of high- and low-fidelity simulations on $32$ cores. 

\paragraph{Estimating the variance}
The reference variance, obtained using the same setup as for the expectation, is $\hat{\sigma}^2_{\text{ref}} = 0.01716$.
We show our results in Table~\ref{tab:mean_var_8d_0_3}.
As for the mean estimator, the MFMC variance estimator is more accurate than the standard MC estimator for the same budget for all investigated cases.
\begin{table}[ht]
\centering
\begin{tabular}{|c|c|c|}
\hline
	$p$ & $\widehat{Var}^{MC}$ & $\widehat{Var}^{MFMC,1}$\bigstrut[t]\\
\hline
	5,000 & 0.02074 & 0.01719 \\
\hline
	10,000 & 0.01771 & 0.01716 \\
\hline
	50,000 & 0.01754 & 0.01715 \\
\hline
\end{tabular}
\caption{The estimated variance of the output of interest for the Cyclone Base Case with eight inputs. The superscript $MFMC,1$ refers to estimators obtained from MFMC using the sparse grid interpolation low-fidelity model. The reference variance is $\hat{\sigma}^2_{\text{ref}} = 0.01716$. 
\label{tab:mean_var_8d_0_3}}
\end{table}

\subsection{Turbulence suppression by energetic particles} \label{subsec:jet}

This scenario is inspired by \cite{Si18,Si19}, which studied the effect of energetic particles such as Neutral Beam Injection (NBI) fast deuterium and Ion Cyclotron Resonance Heating (ICRH) Helium-3 in suppressing ITG turbulence in the Joint European Torus (JET) tokamak. 
It was shown numerically in \cite{Si18,Si19} that the turbulence suppression observed in tokamak experiments can be explained via the combination of two distinct effects: (i) a quasi-linear wave-particle resonance interaction between fast ions and the bulk ITG-driven micro-instabilities and (ii) a nonlinear wave-wave interaction between marginally stable energetic particle-driven modes and ITG turbulence. 
Simulations supported by theoretical results have shown that the former effect takes place when the drift-frequency of the supra-thermal particles gets close to the frequency of the underlying ITG, thus allowing a free energy redistribution. 
Experimental evidence showing signatures of improved plasma confinement due to this quasi-linear effect have been observed in tokamaks \cite{Si21, Si18, Si19}, and predicted also for optimized stellarator devices \cite{W7X}.
The second stabilization effect involves a nonlinear wave-wave coupling between ITG turbulence and marginally stable electromagnetic modes driven by fast ions, which depletes the ITG turbulent drive. 
In strong electromagnetic regimes, these nonlinearly excited fast ion modes act as a catalyst of energy into axisymmetric perturbations called zonal flows, which further reduce turbulent transport \cite{Si_EM2,Si_EM1}.

\begin{figure}[ht]
    \centering
    \includegraphics[width=0.8\textwidth]{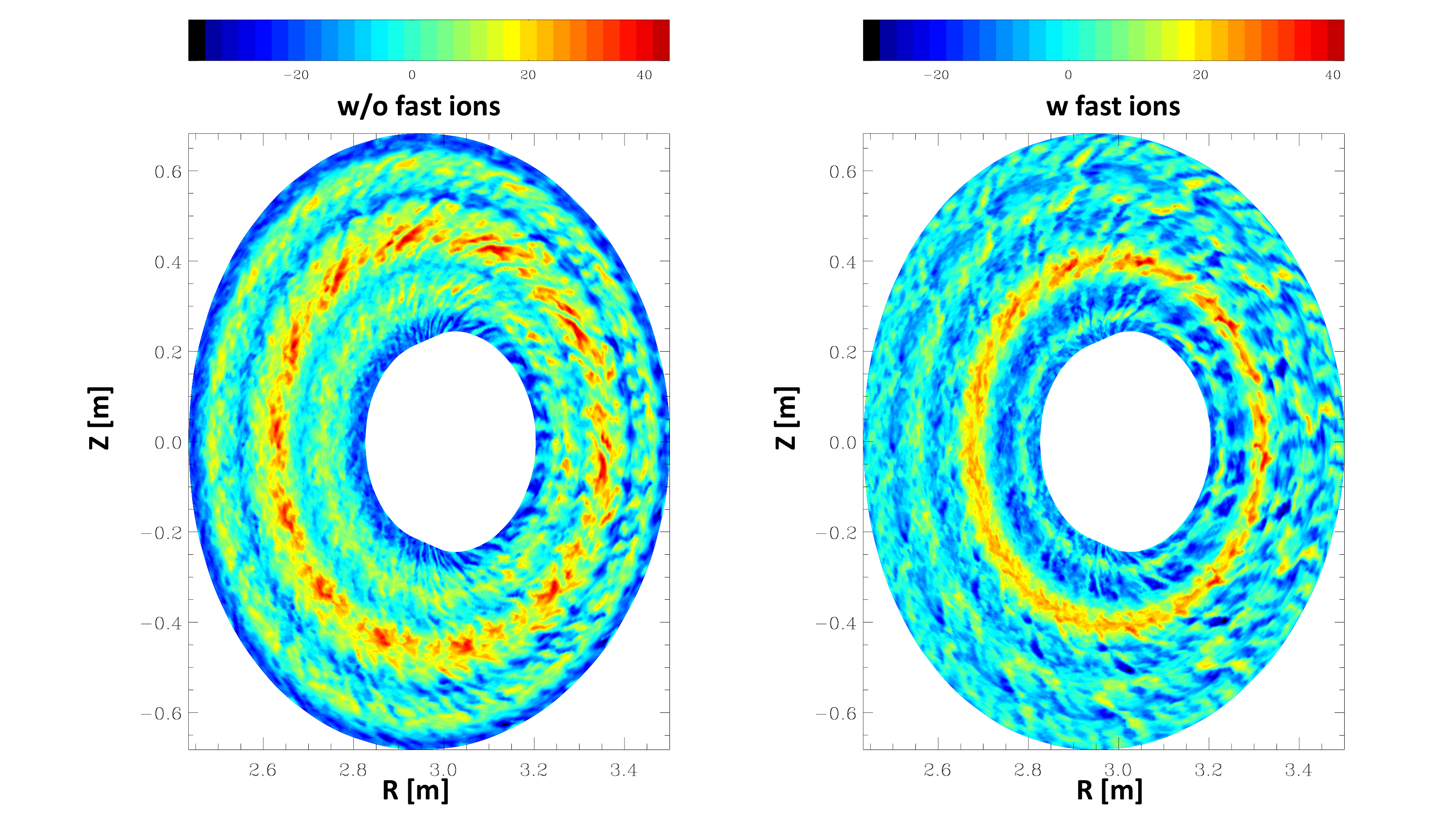}
    \caption{Poloidal cut (at a selected time step) of the density fluctuations in two nonlinear simulations without (left) and with (right) fast ions.
When using fast ions, the streamers are less elongated and do not travel outward as in the case without fast ions.
Therefore, there is less radial transport and hence more turbulence stabilisation in the simulation with fast ions.}
    \label{fig:poloidal_cut}
\end{figure}

To emphasize the effect of fast ions in stabilizing plasma turbulence, we show in Figure \ref{fig:poloidal_cut} a poloidal cut of the density fluctuations from two nonlinear simulations.
The left panel shows the case without fast ions and the right panel the case when fast ions are injected into the plasma.
Using fast ions leads to less elongated streamers and therefore less radial transport, as a result of the combined effect of reduced turbulent drive and increased zonal flow activity, meaning more turbulence stabilization than in the case without fast ions.
Besides the two types of energetic particles, NBI fast deuterium and ICRH $^3$He, and the other two usual particle species in a plasma, deuterium ions and electrons, we also consider Carbon impurities to mimic JET-like C-Wall plasma conditions.
This scenario therefore has a total of five particle species.

The study of turbulence suppression by energetic particles is therefore a problem of high practical relevance in the fusion community, since both experiments and simulations suggest that supra-thermal particles generated via external heating schemes might lead to a significant improvement of the fusion output and overall plasma performances.
And since a potentially large number of parameters characterizing the properties of the magnetic geometry, the bulk species, and the energetic particles are typically affected by e.g., noise in the experimental measurements, it is paramount to perform numerical simulations using an uncertainty propagation framework.

\paragraph{Setup}
To discretize the $5$D gyrokinetic state space, we use $21$ Fourier modes in the radial ($x$) direction and $24$ points along the field line in the $z$ direction.
In velocity space, we employ $32$ equidistant symmetric parallel velocity grid points and $16$ Gauss-Laguerre distributed magnetic moment points. 
This gives a total of $258,048$ degrees of freedom.
The magnetic geometry is characterized by the analytical Miller equilibrium \cite{Mi98}.

The {\sc Gene} simulations were performed using $240$ cores on eight Intel Xeon E5-2697 nodes of the CoolMUC-2 Linux Cluster. 
The low-fidelity models were simulated on an Intel Core i5-8250U CPU running at $1.60$ GHz.
As in the first scenario, all calculations were performed in double precision arithmetic.

The parameters of interest in this problem are the ones associated to the underlying particle species.
Therefore, in our simulations, we consider $12$ uncertain parameters characterizing the main ions, electrons, fast deuterium and $^3$He, and two additional parameters associated to the magnetic geometry, i.e., the safety factor, $q$, and the magnetic shear, $\hat{s}$.
This gives a total of $14$ uncertain parameters.
We model the uncertain parameters as uniform random variables with $25\%$ bounds around their nominal values as shown in Table~\ref{tab:param_14d}.
Throughout our simulations, the impurity parameters are kept fixed to their nominal values.
Quasi-neutrality is ensured by prescribing the values for the density of the deuterium ions and the logarithmic density gradient of electrons in terms of the values of the density and logarithmic density gradients of the other four species, respectively.
We perform our simulations at a fixed wave number, $k_y \rho_s = 0.5$, which is known to be the most unstable ITG eigenmode in flux-tube simulations \cite{Si18}. 
\begin{table}[ht]
    \centering
    \begin{tabular}{|c|c|c|c|}
        \hline
         & parameter & symbol & probability distribution  \\
         \hline
    	$\theta_{1}$ &safety factor & $q$ & $\mathcal{U}(1.3230, 2.1705)$ \\
        \hline        
        $\theta_{2}$ &magnetic shear & $\hat{s}$ & $\mathcal{U}(0.3920, 0.6533)$ \\
        \hline
        $\theta_3$ &ion log temperature gradient & $\omega_{T_i}$ & $\mathcal{U}(3.4230, 5.7050)$\\
        \hline
         $\theta_4$ & ion log density gradient & $\omega_{n_i}$ & $\mathcal{U}(0.0047, 0.0078)$\\
        \hline
        $\theta_5$ &ion temperature & $T_i$ & $\mathcal{U}(0.7500, 1.2500)$ \\
        \hline
        $\theta_6$ &fast deuterium log temperature gradient & $\omega_{T_D}$& $\mathcal{U}(0.7742, 1.2903)$  \\ 
         \hline
        $\theta_7$ &fast deuterium log density gradient & $\omega_{n_D}$ & $\mathcal{U}(3.5413, 5.9022)$ \\ 
         \hline
        $\theta_8$ &fast deuterium density & $n_D$ & $\mathcal{U}(0.0450, 0.0750)$  \\ 
         \hline
        $\theta_9$ &fast deuterium temperature & $T_D$& $\mathcal{U}(7.3500, 12.2500)$  \\ 
         \hline
        $\theta_{10}$ &Helium-3 log temperature gradient & $\omega_{T_{\ce{^{3} He}}}$ & $\mathcal{U}(5.5543, 9.2573)$  \\ 
         \hline
         $\theta_{11}$ &Helium-3 log density gradient & $\omega_{n_{\ce{^{3} He}}}$& $\mathcal{U}(0.3770, 0.6283 )$  \\ 
         \hline
        $\theta_{12}$ &Helium-3 density & $n_{\ce{^{3} He}}$ & $\mathcal{U}(0.0525, 0.0875)$  \\ 
         \hline 
        $\theta_{13}$ & Helium-3 temperature & $T_{\ce{^{3} He}}$ & $\mathcal{U}(9.0000, 15.0000)$ \\ 
         \hline

    	$\theta_{14}$ &electron log temperature gradient & $\omega_{T_e}$ & $\mathcal{U}(1.6695, 2.7825)$\\
    	\hline

    \end{tabular}
    \caption{Input parameters and their distributions used in the scenario of studying turbulence suppression with 14 uncertain inputs. By $\mathcal{U}(a, b) $,  we denote a uniform distribution on $[a, b]$.}
    \label{tab:param_14d}
\end{table}

\paragraph{Data-driven low-fidelity models with full and reduced stochastic dimension}
The low-fidelity models are constructed using sensitivity-driven dimension-adaptive sparse grid interpolation.
Initially, we compute a sensitivity-driven low-fidelity model using all $14$ parameters of Table \ref{tab:param_14d}; this will be the first low-fidelity model, $f^{(1)}$, in the multi-fidelity hierarchy.
We prescribe tolerances $\bm{\tau} = 10^{-6} \cdot \bm{1}_{15}$ and maximum level $L_{\mathrm{max}} = 20$. 
With this setup, we need $96$ {\sc Gene} evaluations to construct the $14$-dimensional low-fidelity model.

\begin{figure}[ht]
    \centering
    \includegraphics[width=0.8\textwidth]{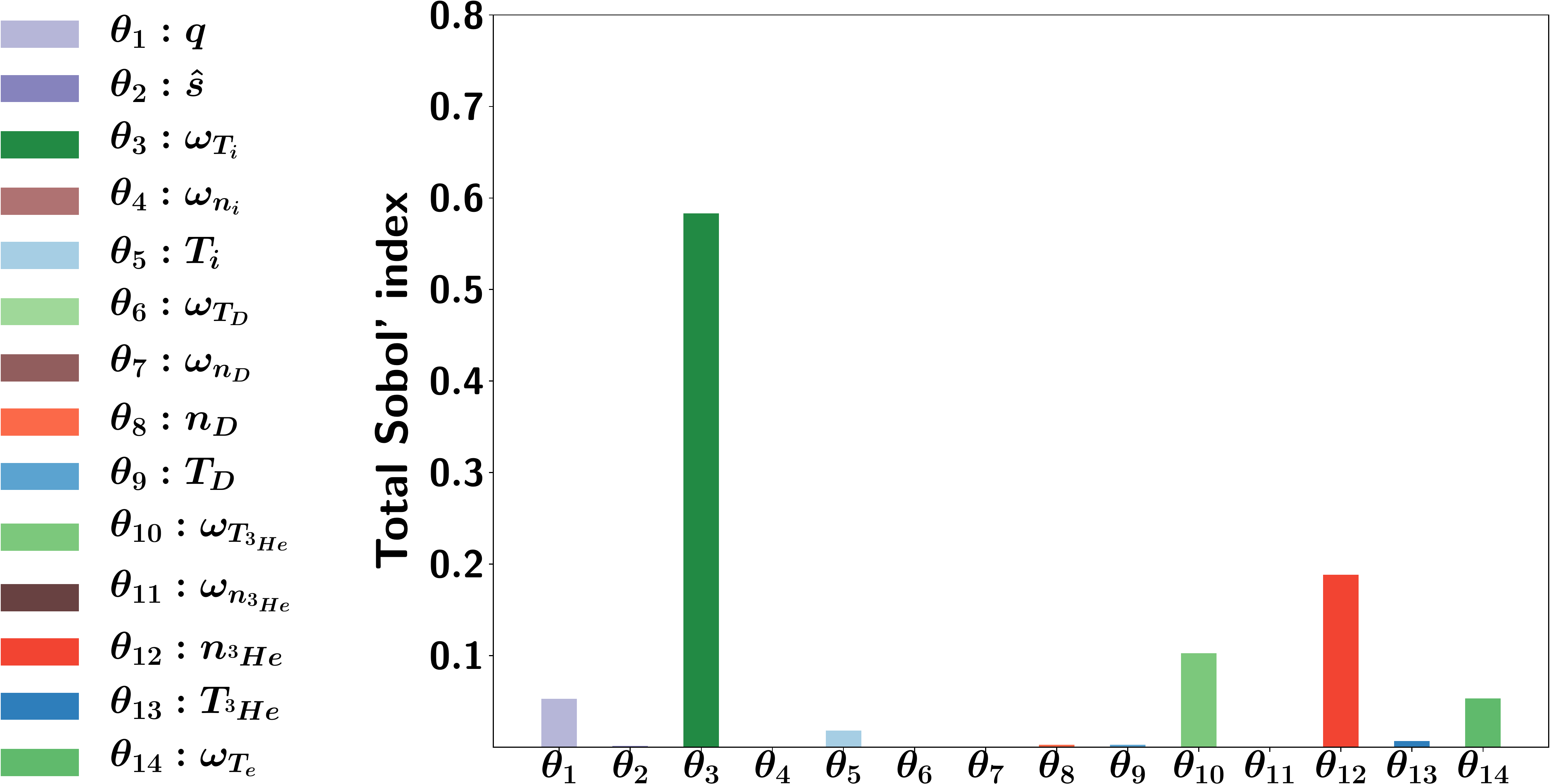}
    \caption{Total Sobol' indices of the 14 stochastic inputs from Table \ref{tab:param_14d} computed by sensitivity-driven dimension-adaptive sparse grid interpolation.}
    \label{fig:sobol}
\end{figure}

We exploit the sensitivity information provided by $f^{(1)}$ and construct reduced-dimension low-fidelity models as well.
To this end, we compute the associated total Sobol' indices, which are visualized in Figure \ref{fig:sobol}.
We observe that at most six parameters ($\theta_1$, $\theta_3$, $\theta_5$, $\theta_{10}$, $\theta_{11}$ and $\theta_{14}$) are important in this scenario; the total Sobol' indices of the remaining eight parameters fall below $0,001$.
In addition, the total indices corresponding to $\theta_2, \theta_4, \theta_6$ and $\theta_7$ are in $O(10^{-4})$ or smaller.
Therefore, the values of the total Sobol' indices suggest decreasing the stochastic dimension from $14$ to nine or even down to six.
For a more comprehensive overview, we consider two reduced-dimension low-fidelity models: $f^{(2)}$, which includes the nine most important parameters and $f^{(3)}$, which includes the six most important stochastic parameters.
We note that the values of the total Sobol' indices in Figure \ref{fig:sobol} are consistent with what is expected from a physics perspective: since we have ITG-driven micro-turbulence, the main ions' logarithmic temperature gradient is expected to be the most important parameter. 
Additionally, for the considered parameter ranges in Table \ref{tab:param_14d}, the Helium$-3$ logarithmic temperature gradient and its density are expected to be important as well \cite{Si18, Si19, FDJ21}.

Tables~\ref{tab:lo_fi_14d} and \ref{tab:lo_fi_14d_2} report the single-core runtimes, correlation coefficients and variances of the high-fidelity model $f^{(0)}$ and the three low-fidelity models $f^{(1)}, f^{(2)}$ and $f^{(3)}$ with their respective inputs.
Since the evaluation of the correlation coefficient and of the variance of the high-fidelity model entails high-fidelity evaluations, it is important to keep the associated cost as low as possible.
To this end, we evaluate the aforementioned quantities using only $50$ MC samples generated independently in an offline step.
Additionally, we also consider the case in which we generate $50$ MC samples to estimate this quantities and re-use these $50$ samples in the MFMC estimators.
In other words, in the latter case we bias the MFMC estimators, the trade-off being that we do not require additional high-fidelity evaluations to estimate the correlation coefficient and the high-fidelity variance.
To distinguish between the two strategies, we will use in the following the label ``re-use" for the latter case.
We see that all low-fidelity models have a high correlation with the high-fidelity model, showing that even the reduced-dimension models will be useful for MFMC.
This is in contrast with traditional single-fidelity approaches in which the full set of uncertain parameters is used.
At the same time, the runtime of the reduced-dimension models is much lower than that of the $14$-dimensional models. 
The six-dimensional model $f^{(3)}$ is $186$ times faster to evaluate than the 14-dimensional low-fidelity model $f^{(1)}$ and about $1.8 \times 10^5$ times faster than the high-fidelity model $f^{(0)}$. 
\begin{table}[ht]
    \centering
    \begin{tabular}{|c|c|c|c|}
         \hline
         $f^{(j)}$ & $d_j$ & $\boldsymbol{\theta_j}$ & $w_j~[sec]$ \bigstrut[t] \\
         \hline
         $f^{(0)}$  & 14D & $\{\theta_1, \theta_2, \theta_3, \theta_4, \theta_5, \theta_6, \theta_7, \theta_8, \theta_9, \theta_{10}, \theta_{12}, \theta_{13}, \theta_{14}\}$ & 11574.8697\bigstrut[t]\\
         \hline
         $f^{(1)}$ & 14D & $\{\theta_1, \theta_2, \theta_3, \theta_4, \theta_5, \theta_6, \theta_7, \theta_8, \theta_9, \theta_{10}, \theta_{12}, \theta_{13}, \theta_{14}\}$ & 11.6724 \bigstrut[t]\\
         \hline
         $f^{(2)}$ & 9D & $\{\theta_1, \theta_3, \theta_5, \theta_8, \theta_9, \theta_{10}, \theta_{12}, \theta_{13}, \theta_{14}\}$ & 0.3838 \bigstrut[t]\\
         \hline
         $f^{(3)}$ & 6D & $\{\theta_1, \theta_3, \theta_5, \theta_{10}, \theta_{12}, \theta_{14}\}$ & 0.0627 \bigstrut[t]\\
         \hline
    \end{tabular}
    \caption{Inputs of low-fidelity models and their runtimes for the turbulence suppression scenario.}
    \label{tab:lo_fi_14d}
\end{table}

\begin{table}[ht]
    \centering
    \begin{tabular}{|c|c|c|c|c|}
         \hline
         $f^{(j)}$ & $\rho_i$ & $\rho_i$ re-use & $\sigma^2_i$ & $\sigma^2_i$ re-use\bigstrut[t] \\
         \hline
         $f^{(0)}$ & 1.0000 & 1.0000 & 0.004379 & 0.002931 \bigstrut[t]\\
         \hline
         $f^{(1)}$ & 0.9978 & 0.9967 & 0.004450 & 0.002897 \bigstrut[t]\\
         \hline
         $f^{(2)}$ & 0.9977 & 0.9963 & 0.004445 & 0.002885 \bigstrut[t]\\
         \hline
         $f^{(3)}$ & 0.9929 & 0.9875 & 0.004415 & 0.002899 \bigstrut[t]\\
         \hline
    \end{tabular}
    \caption{Correlation coefficients and variances of the low-fidelity models for the turbulence suppression scenario estimated from $50$ independent samples and $50$ samples which are also used in MFMC.}
    \label{tab:lo_fi_14d_2}
\end{table}

\paragraph{Estimating the expectation}
We compare standard MC and MFMC for the expectation estimation for budgets $p \in \{5 \times 10^4, 10^5, 5 \times 10^5, 10^6\}$ seconds, ensuring that at least one high-fidelity model evaluation is within the budget.
We consider two MFMC estimators: one estimator that uses only the $14$-dimensional low-fidelity model and another estimator that uses all low-fidelity models. 
In addition, for each estimator we consider the case in which the correlation coefficient and high-fidelity model's variance were computed from separate high-fidelity evaluations and also the case in which the samples used to estimate the two quantities are re-used in MFMC.
The reference expectation and variance were computed using MFMC with a budget $p_{\text{ref}} = 10^7$ seconds.
We obtain $\hat{\mu}_{\text{ref}} = 0.1972$.

\begin{figure}[ht]
\centerfloat
\begin{minipage}[ht]{0.5\textwidth}
\includegraphics[width=\textwidth]{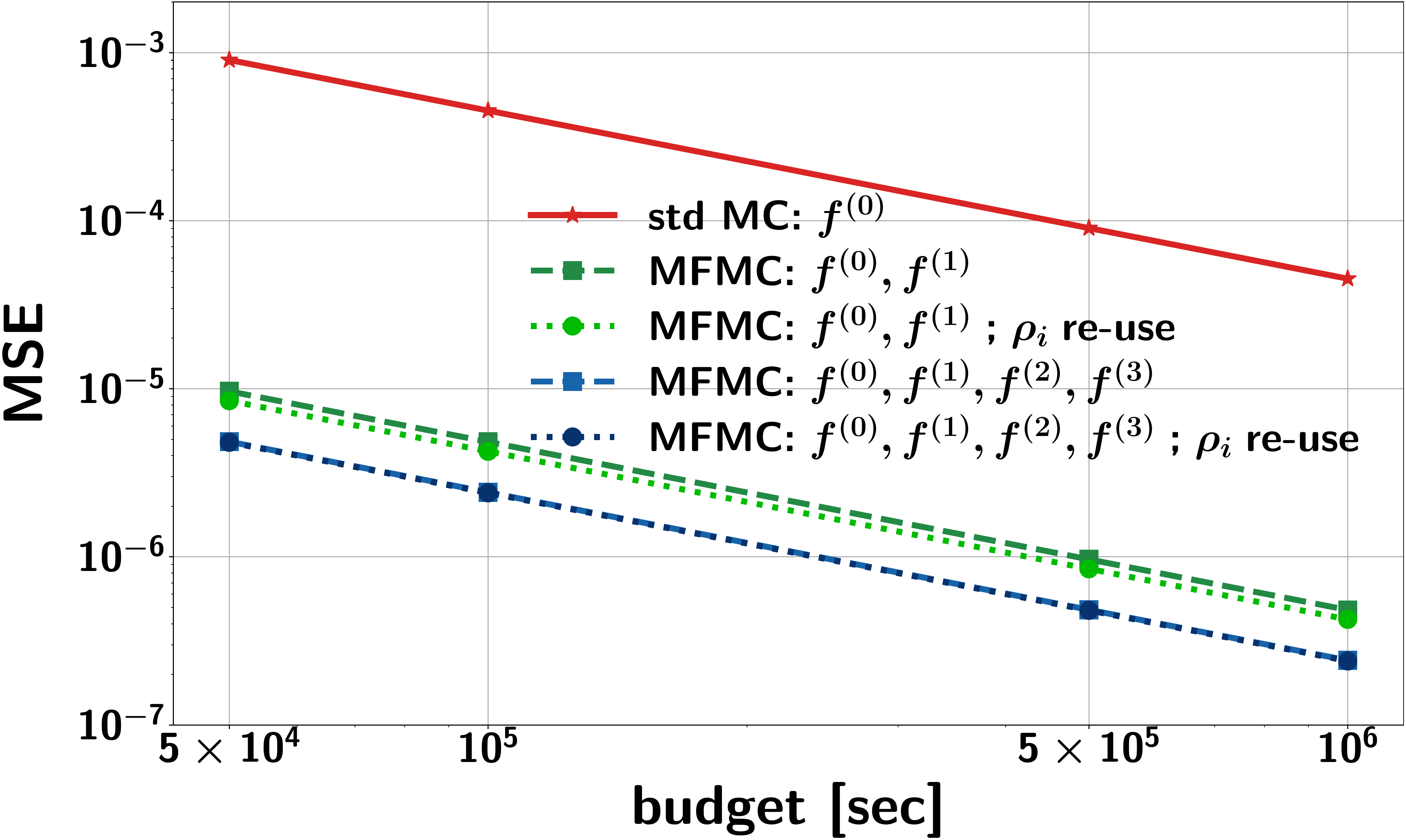}
\end{minipage}
\begin{minipage}[ht]{0.5\textwidth}
\includegraphics[width=\textwidth]{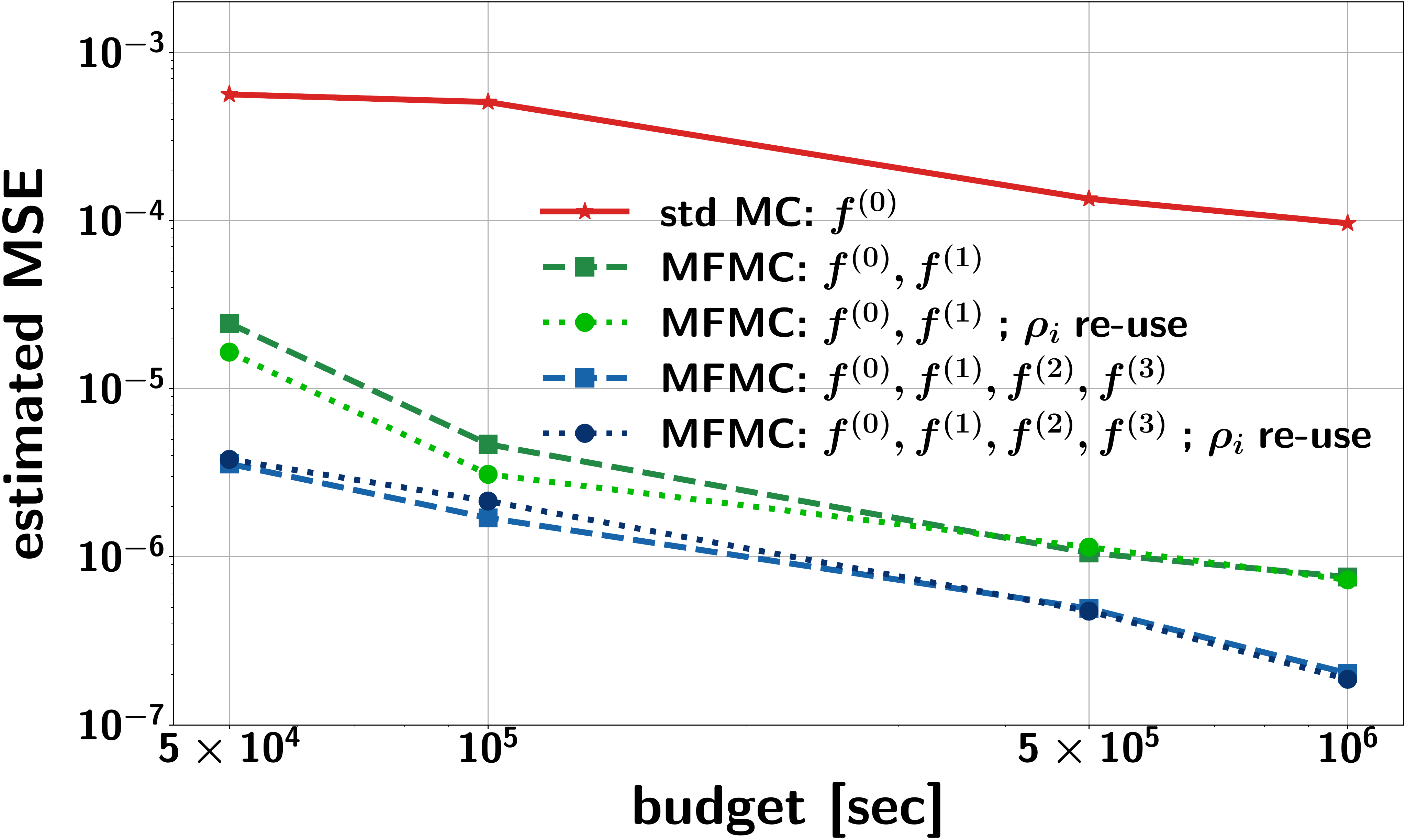}
\end{minipage}
\caption{Analytical MSE (left) and estimated MSE (right) of standard MC and MFMC for the turbulence suppression scenario.}
\label{fig:mse_14d}
\end{figure}

We depict the analytic (left) and the estimated MSE obtained using \eqref{eq:mse_replicates} with $N = 20$ replicates in Figure \ref{fig:mse_14d}.
We observe that the MFMC estimator that uses the high-fidelity model and the $14$-dimensional low-fidelity model is about two orders of magnitude more efficient than standard MC.
Adding the computationally cheap reduced-dimension models leads to an improvement of about a factor $2$ compared to the MFMC with only the full-dimension low-fidelity model.
We see therefore that adding the reduced-dimension low-fidelity models into our data-driven multi-fidelity approach leads to more accurate results.
In addition, we see that reusing the samples used to estimate the correlation coefficient and high-fidelity model's variance has an insignificant effect on the MSE of the MFMC estimators.
\paragraph{Speedup} The obtained variance reduction translates in the following speedup.
To obtain the smallest MSE, i.e., the MSE corresponding to the MFMC estimator with all three low-fidelity models with budget $p = 10^6$ seconds, we need around one hour of high- and low-fidelity simulations on $240$ cores.
The MFMC estimator with only the full-dimension low-fidelity model requires about two hours of high- and low-fidelity simulations on $240$ cores to yield the same MSE.
Finally, for the same task, the standard MC estimator requires a budget of approximately $165,630,360$ seconds on a single core.
To perform the corresponding high-fidelity simulations on $240$ cores requires a total of about eight days for the standard MC estimator.
Thus, our MFMC approach reduces the total runtime on $240$ cores from eight days to one hour when using the estimator that considers both full- and reduced-dimension low-fidelity models.

\paragraph{Estimating the variance} In Table~\ref{tab:var_14d}, we show the variance estimates obtained with MC and MFMC. 
The reference variance is $\hat{\sigma}^2_{\text{ref}} = 0.00397$, computed using the same procedure as for the expectation reference. 
The MFMC estimates are more accurate than the standard MC estimates and adding the dimension-reduced models improves the accuracy of the estimates further.
We observe that between the two MFMC estimators, the most accurate is the MFMC estimator that considers all low-fidelity models.
We therefore see that adding reduced-dimension low-fidelity models improves MFMC variance estimates in this example.
Finally, as we observed for the expected value, we see that reusing the samples used to estimate the correlation coefficient and high-fidelity model's variance has an insignificant effect on the MFMC variance estimators.
\begin{table}[ht]
\centering
\begin{tabular}{|c|c|c|c|c|c|}
\hline
	$p$& $\widehat{\Var}^{MC}$ & $\widehat{\Var}^{MFMC, 1}$ & \shortstack{$\widehat{\Var}^{MFMC, 1}$ \\ $\rho_i$ re-use} & $\widehat{\Var}^{MFMC, 3}$ & \shortstack{$\widehat{\Var}^{MFMC, 3}$ \\ $\rho_i$ re-use} \bigstrut[t]\\
\hline
	$5 \times 10^4$ & 0.00274 & 0.00405 & 0.00420 & 0.00405 & 0.00406 \bigstrut[t]\\
\hline
	$1 \times 10^5$ & 0.00372 & 0.00401 & 0.00404 & 0.00400 &  0.00400 \bigstrut[t]\\
\hline
	$5 \times 10^5$ & 0.00392 & 0.00404 & 0.00403 & 0.00399 & 0.00399 \bigstrut[t]\\
\hline
	$1 \times 10^6$ & 0.00407 & 0.00400 & 0.00401 & 0.00398 & 0.00398 \bigstrut[t]\\
\hline
\end{tabular}
\caption{The estimated variance of the output of interest in the turbulence suppression scenario. The superscripts $MFMC, 1$ and $MFMC, 3$ refer to estimators obtained from MFMC using one ($f^{(1)}$) and three ($f^{(1)}, f^{(2)}, f^{(3)}$) low-fidelity models, respectively. The reference variance is $\hat{\sigma}^2_{\text{ref}} = 0.00397$.}
\label{tab:var_14d}
\end{table}

%% file: sections/05_conclusions.tex
\section{Conclusions}\label{sec:conclusions}
In the present paper, we demonstrated the feasibility of uncertainty propagation in large-scale gyrokinetic simulations by means of multi-fidelity Monte Carlo sampling together with structure-exploiting data-driven low-fidelity models. 
The high-fidelity model was given by the {\sc Gene} code, while the low-fidelity models were constructed using the sensitivity-driven dimension-adaptive sparse grid interpolation procedure which exploits that in many real-world problems, the intrinsic dimension is smaller than the ambient dimension and that the stochastic inputs are anisotropically coupled.
In addition, we also considered a deep-network-based low-fidelity model in a scenario where we had available a large database of numerical experiments to train the model.
In both of our numerical experiments, the MFMC estimators with data-driven low-fidelity models achieved several orders of magnitude speedups compared to standard MC estimators.
Especially for the second numerical example, which studied the role of supra-thermal particles in suppressing instabilities, the speedups achieved with MFMC with reduced-dimension low-fidelity models allowed to perform a much more comprehensive uncertainty propagation study than a standard, single-fidelity Monte Carlo estimator. 
Thus, the present work opens the door towards studies involving a large number of gyrokinetic simulations under real-world conditions, including uncertainty quantification and optimization.

%% file: sections/acknowledgements.tex
\section*{Acknowledgments}
I.-G.F., A.D.S.~and F.J.~were supported by the Exascale Computing Project (No. 17-SC-20-SC), a collaborative effort of the U.S. Department of Energy Office of Science and the National Nuclear Security Administration.
B.P.~acknowledges support by the National Science Foundation under grants CMMI-1761068 and IIS-1901091.
We gratefully acknowledge the compute and data resources provided by the Leibniz Supercomputing Centre (\texttt{www.lrz.de}).